**Thermal conductivity of intercalation, conversion, and alloying lithium-ion battery electrode materials as function of their state of charge**

*Jungwoo Shin [1,2,†,a], Sanghyeon Kim [1,2,†,b], Hoonkee Park [3], Ho Won Jang [3], David G. Cahill [1,2]\*, Paul V. Braun [1,2,4,5]\**

[1]Department of Materials Science and Engineering, University of Illinois Urbana-Champaign, Urbana, Illinois 61801, USA.

[2]Materials Research Laboratory, University of Illinois Urbana-Champaign, Urbana, Illinois 61801, USA.

[3]Department of Materials Science Engineering, Research Institute of Advanced Materials, Seoul National University, Seoul 08826, South Korea.

[4]Department of Chemistry, University of Illinois Urbana-Champaign, Urbana, Illinois 61801, USA.

[5]Beckman Institute for Advanced Science and Technology, University of Illinois Urbana-Champaign, Urbana, Illinois 61801, USA.

*Corresponding author: d-cahill@illinois.edu or pbraun@illinois.edu.
*Postal address: 201 Materials Science and Engineering Building, 1304 W. Green St. MC 246, Urbana, IL 61801, USA

†J. Shin and S. Kim contributed equally to this work.

[a] Current address: Department of Mechanical Engineering, Massachusetts Institute of Technology, Cambridge, MA 02139, USA.
[b] Current address: Chemical Sciences and Engineering Division, Argonne National Laboratory, Lemont, IL, 60439, USA.

**Abstract**

Upon insertion and extraction of lithium, materials important for electrochemical energy storage can undergo changes in thermal conductivity ($\Lambda$) and elastic modulus ($M$). These changes are attributed to evolution of the intrinsic thermal carrier lifetime and interatomic bonding strength associated with structural transitions of electrode materials with varying degrees of reversibility. Using *in situ* time-domain thermoreflectance (TDTR) and picosecond acoustics, we systemically study $\Lambda$ and $M$ of conversion, intercalation and alloying electrode materials during cycling. The intercalation $V_2O_5$ and $TiO_2$ exhibit non-monotonic reversible $\Lambda$ and $M$ switching up to a factor of 1.8 ($\Lambda$) and 1.5 ($M$) as a function of lithium content. The conversion $Fe_2O_3$ and NiO undergo irreversible decays in $\Lambda$ and $M$ upon the first lithiation. The alloying Sb shows the largest and partially reversible order of the magnitude switching in $\Lambda$ between the delithiated (18 W m$^{-1}$ K$^{-1}$) and lithiated states (<1 W m$^{-1}$ K$^{-1}$). The irreversible $\Lambda$ is attributed to structural degradation and pulverization resulting from substantial volume changes during cycling. These findings provide new understandings of the thermal and mechanical property evolution of electrode materials during cycling of importance for battery design, and also point to pathways for forming materials with thermally switchable properties.

**1. Introduction**

Understanding the thermal conductivity ($\Lambda$) of lithium-ion (Li-ion) battery electrode materials is important because of the critical role temperature and temperature gradients play in the performance, cycle life and safety of Li-ion batteries [1-4]. Electrode materials are a major heat source in Li-ion batteries, heat which originates from exothermic redox reactions, entropic heating and joule heating during cycling [5]. If this heat cannot be dissipated, the result is overheating, which in the worst-case scenario can result in fires and explosions [6]. Thus, in designing, modeling and predicting thermal behavior of Li-ion batteries, understanding the thermal properties of the electrode materials is important [7].

The majority of Li-ion battery research has treated temperature as a macroscopic indicator, focusing on the balance between heat generation within the cell and dissipation out of the cell [7,8]. This cell-level of granularity, however, does not provide information into how the thermal properties of the electrode materials change with cycling, and how these changes impact long-term performance. It is worth noting that the primary change in thermal properties of the electrode is due to changes in the internal nanostructure of the constituent materials comprising the electrodes; no information about these changes is provided by cell-level thermal measurements.

Li-ion battery electrodes exhibit materials specific reversible and irreversible electrochemical reactions with lithium-ions, some of which result in significant changes in thermal conductivity. While Cu or Al current collectors common to most Li-ion batteries have high-thermal-conductivities, in all except a few designs, the majority of the heat is dissipated normal to the current collectors, and thus must flow through the anode and cathode active material, as well as the electrolyte and separator to exit the cell [9]. A degree of lithiation decrease in thermal conductivity of electrode materials has potential to create hot spots within a cell, which may trigger undesirable effects including lithium dendrite growth and thermal decomposition of battery components (e.g. electrolyte) [10,11]. Along with the direct technological impact, measurement of thermal conductivity changes of electrode materials during cycling could provide insights into the physical basis of changes in the properties of electrode materials during cycling.

Electrochemical reactions of electrode materials can be generally classified into three reaction categories: intercalation, conversion, and alloying. Intercalation systems are the most mature, have been shown to exhibit reversible lithium storage for thousands of cycles, and are the basis of most current Li-ion batteries. Conversion and alloying reactions offer potentially higher energy densities than intercalation systems, however, so far at the cost of cycle life [12]. A few studies are present in the literature on the thermal and elastic properties of intercalation systems including $LiCoO_2$ [13], $MoS_2$ [14,15] and black phosphorus [16] as a function of state of charge (SOC) and cycle number, while prior to the work here, $\Lambda$ of emerging high capacity conversion and alloying electrode materials had yet to be investigated. Here, we present a systematic study of $\Lambda$ and longitudinal modulus ($M$) evolution of five lithium storage materials cross-cutting all three lithiation/delithiation processes: intercalation ($TiO_2$ and $V_2O_5$), conversion ($Fe_2O_3$ and $NiO$) and alloying (Sb). These electrodes are chosen to represent these three classes of materials, with small (< 10%) to large (~ 150%) volume changes during lithiation/delithiation, and a range of reversible and irreversible structural and electronic transitions [17]. We show $\Lambda$ of these materials changes up to an order of magnitude with SOC and cycle number with varying degrees of reversibility. Our comparative study of these electrode materials provides insights on the relationship between electrochemical reaction mechanisms and their changes in $\Lambda$ and $M$, and thermal and elastic data that will hopefully be useful to the energy storage community.

## 2. Results and Discussion

Accurate analysis of thermal transport in a nanoscale thin-films requires a precise design and control of thermal penetration depth to confines thermal waves in a limited region. We apply *in situ* TDTR to measure $\Lambda$ of electrode materials in a sub-micron thin-film geometry (See methods and Supporting Figure S1-3) [18]. Schematics of *in situ* TDTR measurement of electrode materials are shown in liquid (Figure 1a) or solid-state (Figure 1b) electrochemical cells. The liquid cell is used for intercalation and conversion electrodes ($TiO_2$, $V_2O_5$, $Fe_2O_3$ and NiO on $Al/SiO_2$/sapphire). As intercalation and conversion electrodes exhibit small or moderate volume changes ($\leq$ 100%) during cycling, they can be electrochemically cycled in a liquid cell without delaminating. The alloying electrode (Sb on Au/Al/SiO$_2$/sapphire) undergoes immediate delamination in a liquid cell during the first cycle because of a large volume change > 100% during lithiation. To prevent delamination, we employ an all-solid-state cell and apply ~30 MPa of pressure. Details for the design and assembly of liquid and solid-state cells are described in Supporting Figure S4-5. During lithiation and delithiation, we measure $\Lambda$ changes of electrode materials as a function of state of charge and cycle number. As electrode materials with different lithium storage mechanisms undergo a series of phase transitions with varying degrees of reversibility, they show different trends in thermal conductivity change. Figure 1c-f shows representative thermal conductivity curves of intercalation (TiO$_2$), conversion (Fe$_2$O$_3$) and alloying (Sb) electrode materials during the first lithiation/delithiation cycle. Thermal conductivity at the initial state ($\Lambda_0$), the lowest state ($\Lambda_{low}$) and after completing $n^{th}$ cycle ($\Lambda_n$) are recorded and compared. Figure 1f shows an example of measured and fitted TDTR curves for Sb and lithiated Sb. TDTR also generates and detects acoustic echoes within the electrode materials. During cycling, volume and density changes associated uptake of lithium ions result in a shift in this picosecond acoustic signal that is captured by TDTR (Figure 1f, inset).

## 2.1. Intercalation electrode materials

Intercalation materials exhibit fast and reversible lithium insertion/desertion kinetics. Previous reports have shown reversible $\Lambda$ switching of intercalation electrode materials during lithiation/delithiation. Li$_x$CoO$_2$, Li$_x$MoS$_2$ and Li$_x$P are found to exhibit $\Lambda$ switching contrasts ($\Lambda_{high}/\Lambda_{low}$) of 1.5 – 2.0 [13,14,16]. These intercalation electrode materials consist of 2-D octahedra sheets (CoO$_6$, MoS$_6$) or six-membered phosphorous ring (P$_6$) hosting energetically favorable sites for lithium-ions. Insertion of lithium-ions into these guest sites triggers a phase transition and in various ways can reduce $\Lambda$. To expand on these systems, we choose the prototypical intercalation electrode materials $TiO_2$ (anatase) and $V_2O_5$. $TiO_2$ is tetragonal and consists of edge-sharing $TiO_6$ octahedra chains forming 1-D vacancy channels for lithium-ion storage and transport [19]. $V_2O_5$ consists of 2-D sheets of $VO_5$ square pyramids that are held together by van der Waals (vdW) force and provide 2-D channels for lithium-ion storage and transport.

$TiO_2$ and $V_2O_5$ exhibit theoretical volume expansions of less than 6% upon lithium intercalation [20,21]. Using cross-sectional scanning electron microscopy (SEM) we compare the thicknesses of $TiO_2$ and $V_2O_5$ before and after the lithiation (Supporting Figure S6). We find the volume expansion of $TiO_2$ and $V_2O_5$ as 16% and 4%, respectively. The overall lithiation-delithiation processes with lithium-ions for $TiO_2$ and $V_2O_5$ are:

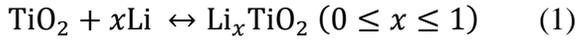
$$TiO_2 + x\text{Li} \leftrightarrow Li_xTiO_2 \ (0 \leq x \leq 1) \quad (1)$$

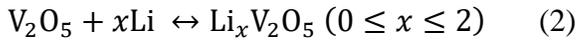
$$V_2O_5 + x\text{Li} \leftrightarrow Li_xV_2O_5 \ (0 \leq x \leq 2) \quad (2)$$

For $TiO_2$, intercalated lithium-ions within the 1-D vacancy channels form $LiO_6$ octahedra with oxygen atoms of the $TiO_6$ chains. As a result, the O-Ti-O bond angle octahedra changes with the inserted lithium-ions, resulting in reversible phase transition between tetragonal α-$TiO_2$ ($x <$ 0.1) and orthorhombic β-$Li_xTiO_2$ (0.1 $< x <$ 0.6) phases [22,23]. Intercalated lithium-ions between 2-D vdW sheets of $V_2O_5$ distort and the 2-D sheet of $VO_5$ pyramids [24], triggering a series of phase transitions between α ($x <$ 0.1), ε (0.35 $< x <$ 0.7), δ ($x \leq$ 1) and γ (1 $< x <$ 2) phases [25]. We stop at $x \approx 2$ since further insertion of lithium-ions can trigger a transition to ω-phase (2 $< x <$ 3), which appears to be irreversible [26].

We measured $\Lambda$ and $M$ of $Li_xTiO_2$ and $Li_xV_2O_5$ as a function of the lithiation state, $x$, using *in situ* TDTR and picosecond acoustics. Prior to TDTR measurement, we confirm the metal/oxygen stoichiometries of $TiO_2$ and $V_2O_5$ using Rutherford backscattering spectrometry (RBS, Supporting Figure S7). We perform X-ray diffraction (XRD, Supporting Figure S8) to confirm the crystallographic orientation and phase transitions of $TiO_2$ and $V_2O_5$. As $TiO_2$ and $V_2O_5$ films are (001) textured, $M \approx C_{33}$ of $TiO_2$ and $V_2O_5$ (Supporting Figure S9). $M$ is estimated as $M(x) = \rho(x)v(x)^2$ where $\rho$ is the density and $v$ is the longitudinal speed of sound of the electrode material measured by picosecond acoustics at $x$ (Supplementary Figure S9). As $TiO_2$ and $V_2O_5$ films are (001) textured, $M \approx C_{33}$ of $TiO_2$ and $V_2O_5$. $x$ of $Li_xTiO_2$ was estimated based on the theoretical electrochemical capacity.

Figure 2a shows $\Lambda$ and $M$ of $Li_xTiO_2$ with respect to $x$. During lithiation, $M$ increased from 164 to 250 GPa, within the range of calculated values of $C_{33} = 185$ and 276 GPa of anatase and Li-titanite [27]. In contrast to the monotonic increase of $M$, we also observe a decrease in $\Lambda$ from 7.7 W m$^{-1}$ K$^{-1}$ ($x = 0$) to 4.4 W m$^{-1}$ K$^{-1}$ ($x = 0.34$) during the first lithiation. During the subsequent delithiation, $\Lambda$ recovers to 8.0 W m$^{-1}$ K$^{-1}$ ($x = 0.1$).

Figure 2b shows $\Lambda$ and electrochemical potential curves of $TiO_2$ over five galvanostatic cycles. During lithiation, initially a small amount of lithium forms a solid solution with $\alpha$-$TiO_2$ [28]. This region corresponds to a small potential slope before the discharge plateau in Figure 2b. $\Lambda$ of $TiO_2$ does not measurably change in this region. As lithium insertion progresses, $TiO_2$ becomes biphasic with coexisting Li-rich ($\beta$-$TiO_2$) and Li-poor ($\alpha$-$TiO_2$) phases [29]. The two-phase equilibrium in this lithiation region is confirmed by the discharge potential plateau at ~1.75 V vs Li/Li$^+$. $\Lambda$ of $TiO_2$ decreases within this region as the Li-rich phase and biphasic boundaries grow as the concentration of lithium increases. For intercalation electrode materials such as $MoS_2$,

Λ decreases could be attributed to phonon scattering with inserted lithium [15]. Because Λ decrease is concentrated at the early stage of lithiation, we speculate that phonon boundary scattering between Li-rich and Li-poor phase is the major reason that Λ is suppressed [13]. Further lithiation has less effect on Λ as the effect of boundary scattering decreases with growth of the low-thermal-conductivity β phase. This region is indicated by a gradual potential slope from 1.75 V vs Li/Li$^+$ to the cut-off voltage. We note, we do not observe a second discharge potential plateau corresponding to the β to γ phase transition because of the thickness of the TiO$_2$ film. Size-dependent lithium-insertion experiments show that lithiation of anatase above $x > 0.6$ (γ-TiO$_2$) is kinetically limited to a few tens of nanometers within a practical cycling rate at room temperature (γ-TiO$_2$ is a poor lithium-ion conductor due to fully occupied vacancy channels) [30].

With each cycle, we observe horizontal shifts in potential curves due to the irreversible electrochemical capacity loss at each cycle. Λ of Li$_x$TiO$_2$ shows a decreasing trend with an increasing number of cycles, probably due to accumulation of defects within the material. However, the Λ contrast between the lithiated and delithiated states remains nearly constant at a 1.6 (7.7 vs. 4.7 W m$^{-1}$ K$^{-1}$ during the 1$^{st}$ cycle, 6.3 vs. 4.0 W m$^{-1}$ K$^{-1}$ during the 5$^{th}$ cycle).

Figure 2c presents Λ and $M$ of Li$_x$V$_2$O$_5$. Compared with a limited lithiation capacity of TiO$_2$ ($x < 0.4$), V$_2$O$_5$ fully lithiates up to $x = 2$. We estimate $x$ in Li$_x$V$_2$O$_5$ based on three characteristic potential plateaus corresponding to the ε ($0.35 < x < 0.7$), δ ($x < 1$) and γ ($1 < x \leq 2$) phases. Due to the presence of 2-D diffusion pathways, V$_2$O$_5$ exhibits higher lithium-ion diffusion coefficients ($D_{Li}$ ~$10^{-11\pm2}$ cm$^2$ s$^{-1}$) [31-33] relative to TiO$_2$ ($D_{Li}$ ~ $10^{-13\pm2}$ cm$^2$ s$^{-1}$) [34,35]. We estimate $M$ of as-prepared V$_2$O$_5$ as 24 GPa from picosecond acoustics (Supporting Information Figure S9). This value is lower than $C_{33}$ of graphite (37 GPa) [36]. Previous experiments show that the speed of sounds and elastic modulus of V$_2$O$_5$ vary significantly with preparation conditions [37,38]. The elastic modulus of $V_2O_5$ is highly sensitive to sample geometry, synthesis condition, stoichiometry, orientation, crystallinity and defects [39]. Ab-initio calculations also show variation in bulk modulus of $V_2O_5$ as (18 – 68 GPa) [40-42]. Similar to $TiO_2$, $M$ gradually increases to 28 GPa during lithiation to $x = 2$. Contrary to the monotonic increase in $M$ with lithiation, we observe a non-linear $\Lambda$ response of $Li_xV_2O_5$, including a $\Lambda$ decrease from 7.2 W m$^{-1}$ K$^{-1}$ ($x = 0$) to 4.7 W m$^{-1}$ K$^{-1}$ ($x = 0.5$) followed by an increase to 5.4 W m$^{-1}$ K$^{-1}$ ($x = 1$). During further lithiation ($x = 2$), $\Lambda$ gradually decreases to 4.8 W m$^{-1}$ K$^{-1}$.

Figure 2d presents $\Lambda$ and electrochemical potential curves of $Li_xV_2O_5$ over five galvanostatic cycles. During the first lithiation process, $V_2O_5$ exhibits three distinct discharge potential plateaus at around 3.0, 2.5 and 2.0 V vs. Li/Li$^+$, respectively. These plateaus are found lower than the discharge plateaus during the subsequent cycling (3.2, 3.0 and 2.1 V vs. Li/Li$^+$) due to large overpotential during the first lithiation process [26]. $\Lambda$ of $V_2O_5$ decreases during the first discharge plateau ($\alpha$ to $\varepsilon$), reaches a local minimum at the end of the first discharge plateau ($\varepsilon$), increases during the second plateau region ($\varepsilon$ to $\delta$) and reaches a local maximum at the start of the third discharge plateau ($\delta$). Then, $\Lambda$ of $V_2O_5$ gradually decreases within the third plateau region ($\delta$ to $\gamma$).

During the subsequent delithiation process (charge), three potential plateaus regions are reduced as some lithium-ions are trapped. The $\Lambda$ valley is also shifted with respect to the potential plateau shift. Specifically, the $\Lambda$ (open circles) of $V_2O_5$ slightly increases within the third charge potential plateau region ($\gamma$ to $\delta$). Then, $\Lambda$ starts to decrease at the beginning of the second charge potential plateau ($\delta$ to $\varepsilon$), reaches a local minimum at the end of the second charge potential plateau ($\varepsilon$) and increases again with the first potential plateau ($\varepsilon$ to $\alpha$). $\Lambda$ during the rest of delithiation processes are shown in Supporting information Figure S10. During the subsequent lithiation/delithiation cycle (2$^{nd}$), we limit the lithiation state up to $x = 1$ (Figure 2d, inset) to confirm the non-monotonic thermal response of V$_2$O$_5$ within two potential plateaus (α to ε and ε to δ). The local Λ minimum is found at the inflection point between the first and second potential plateaus during the lithiation and delithiation processes. During the rest of lithiation/delithiation cycles, the local Λ minima consistently appear. We hypothesize this "thermal conductivity valley" originates from accumulation and relaxation of local disorder and puckering of pyramidal VO$_5$ layers [43].

Λ and elastic modulus of Li$_x$TiO$_2$ and Li$_x$V$_2$O$_5$ show opposite trends with respect to $x$. This behavior is similar to computational results for Li-intercalated graphite and Li$_x$MoS$_2$ where, at low lithium content, randomly intercalated lithium ions scatter phonons but at higher lithium contents eventually increases Λ due to acoustic phonon stiffening [44-46]. With each cycle, we observe horizontal shifts in potential curves due to the irreversible electrochemical capacity loss at each cycle. Λ of intercalation electrode materials shows a decreasing trend with an increasing number of cycles, probably due to accumulation of defects within the material. The electrochemical charge/discharge capacities of TiO$_2$ and V$_2$O$_5$ during *in situ* TDTR measurements are presented in Supporting Figure S11.

## 2.2. Conversion electrode materials

Since the discovery of conversion reactions of metal-oxides with lithium [47], conversion electrodes have been studied as promising Li-ion battery electrode materials due to their high theoretical capacities. Conversion electrode materials typically lithiate via a kinetically slow process in which lithium at the electrode surface diffuse into and react with the bulk, forming a moving reaction front between Li-rich and poor phases [48]. As the reaction front advances into Li-poor phase, the conversion reaction takes place. The conversion reaction involves a complete reconfiguration of metal and oxygen atoms where metal oxide ($MO_x$) is converted into lithium oxide ($Li_2O$) and pure metal phases. After $Li_2O$/metal phases form, lithium diffusion in this region is further suppressed [48].Thus, the overall kinetics is controlled by both diffusion and conversion reaction at the reaction front. During the lithiation/delithiation, conversion electrode materials undergo substantial volume expansion/contraction along with amorphization.

We choose $Fe_2O_3$ and NiO as model conversion systems. Due to slow diffusion kinetics for alloying electrode materials, diffusion coefficients of lithium in $Fe_2O_3$ ($D_{Li} \sim 10^{-14\pm1}$ cm$^2$ s$^{-1}$) [48] and NiO ($D_{Li} \sim 10^{-13\pm1}$ cm$^2$ s$^{-1}$) [49] are orders of magnitude lower than intercalation electrode materials. Both $Fe_2O_3$ and NiO are poor electrical conductors, in both lithiated and delithiated forms. Thus, heat is carried by phonons, and crystallinity plays an important role in thermal conductivity. The overall lithiation processes for $Fe_2O_3$ and NiO are as follows:

$$Fe_2O_3 + 6x Li \leftrightarrow 3x Li_2O + 2x Fe + (1-x) Fe_2O_3 \ (0 \leq x \leq 1) \quad (3)$$

$$NiO + 2x Li \leftrightarrow x Li_2O + x Ni + (1-x) NiO \ (0 \leq x \leq 1) \quad (4)$$

One mole of $Fe_2O_3$ and NiO can take up to six and two moles of Li-ions, respectively. The degree of lithiation $x$ is defined as the maximum molar lithium capacity per one mole of $Fe_2O_3$ and NiO. RBS data suggest that $Fe_2O_3$ and NiO films contain inhomogeneous metal-rich phases ($Fe_3O_4$ and Ni) (Supporting Figure S7). We confirm irreversible amorphization of $Fe_2O_3$ and NiO using *ex-situ* XRD analysis (Supporting Figure S8).

Figure 3a shows $\Lambda$ and the $M$ of $Fe_2O_3$ measured using *in situ* backside TDTR. We measure $M$ of $Fe_2O_3$ electrode as 287 GPa from picosecond acoustic echoes (Supporting Figure S9). This value is larger than the bulk modulus of polycrystalline $Fe_2O_3$ (223 GPa) [50] and slightly smaller than the measured $C_{33} = 299$ GPa of a single crystal α-$Fe_2O_3$ [51]. After the first lithiation ($x$ = 1), the picosecond acoustic echoes diminish along with a significant $\Lambda$ drop from 3.9 W m$^{-1}$ K$^{-1}$ to 0.9 W m$^{-1}$ K$^{-1}$. $\Lambda$ of Fe$_2$O$_3$ does not recover during delithiation. Regardless of the initial $\Lambda$ of Fe$_2$O$_3$ (1.7 ~ 7.6 W m$^{-1}$ K$^{-1}$, depending on annealing conditions. See Supporting Figure S3), $\Lambda$ of Fe$_2$O$_3$ reduces to a similar level (< 1 W m$^{-1}$ K$^{-1}$) after the first lithiation.

Figure 3b shows $\Lambda$ and electrochemical cycling curve of Fe$_2$O$_3$ over five galvanostatic cycles. During the first lithiation, a small potential shoulder at 0.8 V vs Li/Li$^+$ is observed, indicating the formation of a solid solution of the cubic γ-Li$_x$Fe$_2$O$_3$ phase at the early stage of lithiation [52]. In this region, $\Lambda$ of Fe$_2$O$_3$ gradually decreases by 10 %. Further lithiation triggers a biphasic reaction involving a reduction of Fe(III) to Fe(0) and the formation of Li$_2$O. During the biphasic conversion reaction, Fe$_2$O$_3$ shows a distinct potential plateau at 0.75 V vs Li/Li$^+$. In this region, $\Lambda$ of Fe$_2$O$_3$ decreases more rapidly (0.4 < $x$ < 0.8) than the initial $\Lambda$ slope. As Fe$_2$O$_3$ lithiates close to its theoretical capacity, $\Lambda$ shows a sharp drop in a narrow region (0.8 < $x$ < 0.9). Fe$_2$O$_3$ loses about 70% of its $\Lambda$ in this region. Starting with second cycle, Fe$_2$O$_3$ no longer exhibits the discharge plateau at 0.75 V vs Li/Li$^+$, but a gradual potential slope from 1.5 V, followed by a potential shoulder at 0.9 V. This potential slope indicates Fe$_2$O$_3$ becomes amorphous after the first cycle [53].

Figure 3c shows $\Lambda$ and $M$ of NiO. At the initial state ($x$ = 0), NiO exhibits $\Lambda$ = 7.7 W m$^{-1}$ K$^{-1}$. We measure $M$ = 296 GPa estimated from the longitudinal acoustic echoes (Supporting Figure S9). As the NiO electrode is (111) textured, this value is slightly lower than ($C_{11}$+2$C_{12}$+4$C_{44}$)/3 = 310 GPa, measured by a longitudinal wave propagating along the [111] direction of a single-crystal NiO [54]. Upon lithiation to $x$ = 0.71, $\Lambda$ reduces to 3.1 W m$^{-1}$ K$^{-1}$ and $M$ decreases to 190 GPa. As lithiation of NiO triggers amorphization, $M$ at $x$ = 0.71 is equivalent to $C_{11}$ = 190 GPa of an isotropic matrix. $\Lambda$ and $M$ do not recover during delithiation.

Figure 3d shows Λ and potential curves of NiO. During the first lithiation, NiO shows a distinct discharge potential plateau at 0.7 V vs Li/Li$^+$. Similar to Fe$_2$O$_3$, the potential plateau is replaced with a gradual potential slope from 1.5 V vs Li/Li$^+$ and a small potential shoulder at 0.75 V vs Li/Li$^+$ after the first cycle (Figure 3d).

The difference between the discharge curves of Fe$_2$O$_3$ and NiO suggests that the first and the rest of the electrochemical cycles take different lithiation/delithiation pathways. The first lithiation involves irreversible processes. During the first lithiation microcracks, voids, defects and vacancies can be formed along with formation of SEI, breaking of M-O bonds, rearrangement of atoms and even pulverization of the electrode materials [55]. As electrochemical potential is related to the total change in Gibbs free energy Δ$G$, these irreversible processes result in large overpotentials and poor coulombic efficiency.

From the second cycle, microcracks and voids act as facile lithium-ion diffusion pathways. Defects and vacancies provide energetically favorable sites for lithium-ions [56]. Lithiation in a random network of metal-oxide (MO$_x$) polyhedra with defects and vacancies requires a smaller Δ$G$ compared with crystalline counterparts upon conversion reaction with lithium [53]. Therefore, metal-oxide conversion electrodes exhibit improved coulombic efficiencies and less potential hysteresis from the second cycle [57]. Though cracks, voids, defects and amorphization could help to enhance lithiation/delithiation kinetics, they strongly scatter phonons and significantly reduce Λ of conversion electrodes after the first cycle.

As thermal transport is mediated by phonons, Λ of Fe$_2$O$_3$ and NiO decreases with the growth of the amorphous phase. Λ of amorphous materials can be estimated using the atomic density and speed of sounds where the minimum mean-free-path of vibrational modes is reduced to the interatomic distance. With this approach, we have successfully predicted asymptotic Λ of disordered materials including dielectrics and polymers [58,59]. We estimate the minimum lattice thermal conductivity for NiO and Fe$_2$O$_3$ as $\Lambda_{min} \sim 1.4$ W m$^{-1}$ K$^{-1}$ and Li$_2$O as $\Lambda_{min} \sim 1.7$ W m$^{-1}$ K$^{-1}$. While $\Lambda$ of NiO remains above $\Lambda_{min}$, $\Lambda$ of Fe$_2$O$_3$ drops below $\Lambda_{min}$ during lithiation $x > 0.8$.

We note that Fe$_2$O$_3$ exhibits a volume expansion 40% larger than theory while NiO shows a volume expansion similar to that theoretically expected. The excess volume expansion could explain why $\Lambda$ of Fe$_2$O$_3$ drops below $\Lambda_{min}$. We speculate that this additional reduction in $\Lambda$ is extrinsic to the material property and due to the formation of voids, microcracks and low thermal conductivity SEI within the electrode materials.

The electrochemical charge/discharge capacities and coulombic efficiencies of Fe$_2$O$_3$ and NiO during *in situ* TDTR experiments are presented in Supporting Figure S12.

### 2.3. Alloying electrode materials

Alloying electrodes, e.g., Si, Sn and Sb, form binary alloys with up to 4.4 Li-atoms per host atom. Li-sulfur and Li-air electrode materials also fall into this class. As a result, alloying electrode materials exhibit high theoretical capacities, ranging from 660 mA h g$^{-1}$ to 4200 mA h g$^{-1}$, however, at a cost of up to a 300% volume expansion during lithiation [60-63]. The considerable volume changes often result in large irreversible capacities, as well as continuous formation/breakage of SEI and structural degradation (pulverization) during cycling [60]. Among alloying electrode materials, we choose Sb because it undergoes fewer phase transformations than Si and Sn (which exhibit many intermediate phases), and has a moderate lithium diffusion coefficient ($D_{Li} \sim 10^{-9\pm2}$ cm$^2$ s$^{-1}$) [64,65] and a discharge potential plateau above 0.5 V vs Li/Li$^+$, making it compatible with the Al transducer. Upon reaction with lithium, Sb undergoes a reversible phase transition between Sb and Li$_3$Sb as follows:

$$\text{Sb} + x\text{Li} \leftrightarrow \text{Li}_x\text{Sb} \ (0 \leq x \leq 3) \quad (5)$$

The corresponding phase transition between Sb and Li$_3$Sb was confirmed using XRD (Supporting Figure S8). The theoretical volume expansion of Sb is calculated as 130% at the fully lithiated state. Using this solid-state design, a volume expansion ratio of 110% is observed, which is however largely associated with the surface deformation (Supporting Figure S6).

Figure 4a shows $\Lambda$ and $M$ of Li$_x$Sb measured using *in situ* TDTR with the solid-state cell. The lithiation state $x$ is defined by the theoretical capacity ($x = 3$, $Q = 660$ mA h g$^{-1}$), the density and the thickness of the Sb film. First, we measure $\Lambda = 18$ W m$^{-1}$ K$^{-1}$ for pristine Sb using frontside and backside TDTR. Sb is a semimetal with overlapping valence and conduction bands at $E_\text{F}$. We estimate the electronic thermal conductivity $\Lambda_\text{e} = 12$ W m$^{-1}$ K$^{-1}$ using the Wiedemann Franz law and the electrical conductivity measured by four-point probe. The lattice thermal conductivity $\Lambda_\text{L}$ is estimated by subtracting $\Lambda$ with $\Lambda_\text{e}$. The estimated value of $\Lambda_\text{L} = 6$ W m$^{-1}$ K$^{-1}$ agrees with the first principle calculation for crystalline Sb [66].

During the first lithiation process, we observe a dramatic $\Lambda$ decrease from 18 W m$^{-1}$ K$^{-1}$ ($x = 0$) to 0.6 m$^{-1}$ K$^{-1}$ ($x = 2.9$). This markedly large $\Lambda$ change can be primarily attributed to the elimination of the electronic contribution to $\Lambda$ during the semimetal to semiconductor (Li$_3$Sb) transition. A similar degree of $\Lambda$ reduction is predicted for Zn$_3$Sb, GaSb and CdSbAg during semiconductor-to-metal transitions at the melting point [67].

Along with the electronic transition, $\Lambda_\text{L}$ plays an important role in decreasing $\Lambda$ at the lithiated state. Sb alloys show significantly low $\Lambda_\text{L}$ compared with Sb. For example, amorphous Sb$_2$Te$_3$ shows $\Lambda = 0.23$ W m$^{-1}$ K$^{-1}$ [68] and disordered Zn$_4$Sb$_3$ shows $\Lambda = 0.6$ W m$^{-1}$ K$^{-1}$ at 300 K [69], due to the large phonon anharmonicity induced loose coupling of atoms with different masses [70]. A first-principles calculation predicts the suppression of $\Lambda_\text{L}$ in the lithiated state (crystalline

Li$_3$Sb, $\Lambda_L$ = 2.2 W m$^{-1}$ K$^{-1}$) [71]. Further decrease in measured $\Lambda$ below 1 m$^{-1}$ K$^{-1}$ can be attributed to the loss of crystallinity. We estimate the minimum lattice thermal conductivity as $\Lambda_{min}$ = 0.43 W m$^{-1}$ K$^{-1}$ for amorphous Li$_3$Sb. Along with a decrease in $\Lambda$, Sb shows a decreasing trend in $M$ with increasing $x$. We estimate $M$ of Sb ($x$ = 0) as 81 GPa. This value is slightly lower than the measured $C_{11}$ = 99 GPa of single-crystal Sb (Supporting Figure S9) [72]. Upon lithiation to $x$ = 1, $M$ decreases to 56 GPa. $x$ > 1 remains uncertain because further lithiation results in large surface deformation and thickness uncertainty.

Figure 4b shows $\Lambda$ and electrochemical potential curves of Sb during five lithiation/delithiation cycles. First, Sb shows a gradual $\Lambda$ decrease from 18 to 12 W m$^{-1}$ K$^{-1}$ up to $x$ = 0.9, followed by a sharp drop in $\Lambda$ at $x$ ~ 1. $\Lambda$ of Sb decreases to 0.7 W m$^{-1}$ K$^{-1}$ at the fully lithiated state. During the subsequent delithiation, $\Lambda$ of Sb recovers up to 9.4 W m$^{-1}$ K$^{-1}$ with a large $\Lambda$ hysteresis, suggesting that Sb loses about 50% of $\Lambda$ after one full lithiation/delithiation cycle. This thermal conductivity hysteresis can be attributed to the phase transition pathway of Sb. Lithium insertion into Sb forms hexagonal Li$_2$Sb phase, followed by cubic Li$_3$Sb phase. During delithiation, Li$_3$Sb directly transforms to Sb without forming Li$_2$Sb first [73]. During the remaining lithiation/delithiation cycles, $\Lambda$ of Sb gradually decreases to ~ 1.3 W m$^{-1}$ K$^{-1}$, and eventually becomes constant when cycling between the lithiated and delithiated states. During five lithiation/delithiation cycles, Sb exhibits discharge plateaus at around 0.8 – 0.85 V vs Li/Li$^+$ due to the phase transition between the Sb, Li$_2$Sb and Li$_3$Sb phases. The discharge plateaus for Li$_2$Sb and Li$_3$Sb are difficult to distinguish, as they are only separated by ~ 0.05 V. However, we observe a large irreversible capacity and poor coulombic efficiency, which appears as horizontal shifts of potential curves.

This large irreversible $\Lambda$ loss of Sb is not a result of intrinsic electronic and phase transitions, but rather electrode pulverization associated with repeated volume expansion and contraction. Electrode pulverization introduces cracks and pores in the films during cycling [74]. This structural degradation can contribute to capacity fading, electric contact loss and large irreversible capacity of alloying electrode materials. Since a majority of phonons carrying heat in Sb and Li$_3$Sb have a mean free path between 10 nm to 100 nm [66,71], electrode pulverization introduces substantial phonon scattering. Furthermore, the loss of electric contact between the pulverized particles leads to a loss of $\Lambda_e$.

These changes are detrimental to $\Lambda$ of electrode materials because they add boundary and defect scattering, and introduce low-thermal-conductivity phases including SEI and voids. In this regard, we hypothesize that the degree of lithiation (or volume expansion) is directly related to irreversible $\Lambda$. To test our hypothesis, we limit lithium intake of Sb to $Q = 250$ mA h g$^{-1}$ (equivalent to $x = 1.14$) per each lithiation process. After lithiation, we freely delithiated Sb up to 1.5 V vs Li/Li$^+$. Figure 4c shows $\Lambda$ and electrochemical potential curves of Sb with the limited discharge. Both full and limited Sb show a similar trend in $\Lambda$ decrease up to $x \sim 1$, which is about the irreversible capacity of the fully activated Sb at the first cycle. However, further cycling results in a noticeable difference in $\Lambda$ retention and hysteresis. As limiting the discharge avoids formation of Li$_3$Sb, we do not observe noticeable $\Lambda$ hysteresis during lithiation and delithiation. Sb cycled over a limited range shows much higher $\Lambda$ at the delithiated state (10.4 – 4.9 W m$^{-1}$ K$^{-1}$) and lithiated state (8.2 – 3.1 W m$^{-1}$ K$^{-1}$), and shows $\Lambda$ switching ratios $\Lambda_{high}/\Lambda_{low} > 1.5$ from 2$^{nd}$ to 5$^{th}$ cycle, similar to intercalation electrode materials.

## 3. Reversible and Irreversible Thermal Conductivity Trends

From the lithium-ion electrode materials was investigated, several trends emerge. Figure 5 summarizes the $\Lambda$ changes of the electrode materials in this study, $V_2O_5$, $TiO_2$, $NiO$, $Fe_2O_3$ and Sb. We plot $\Lambda$ of each electrode at the initial state ($\Lambda_0$, high thermal conductivity state) and $\Lambda$ contrast, using the lowest $\Lambda$ state ($\Lambda_{low}$) at each cycle in Figure 6a and include previous reports. It is worth noting black phosphorus (BP) could be both an intercalation and conversion electrode material depending on their lithiation. For $x < 0.4$ (in $Li_xP$), BP is an intercalation electrode material showing reversible thermal conductivity switching [16]. However, it is also subject to an alloying reaction upon further lithiation, during which P-P bonds break, and $Li_3P$ forms concurrent with a large volume expansion of ~200% [75,76]. In the alloying reaction regime, $Li_xP$ shows irreversible thermal conductivity loss of up to a factor of five, in both the in-plane and out-of-plane direction. Intercalation and alloying $Li_xP$ are plotted separately.

    Intercalation electrodes, including $V_2O_5$ and $TiO_2$, show reversible $\Lambda$ and elastic modulus switching, ascribed to their reversible crystalline phase transitions during lithiation/delithiation. In all these systems, $\Lambda_{low} > 1$ W m$^{-1}$ K$^{-1}$. On the contrary, conversion electrodes undergo irreversible $\Lambda$ decrease of a factor of two to five during the first lithiation process and no further change in $\Lambda$ during subsequent cycling. Therefore, they show low $\Lambda_{low}$ and small $\Lambda_{low}/\Lambda_0$. Conversion Sb shows the largest changes in thermal conductivity between $\Lambda_0 = 18$ W m$^{-1}$ K$^{-1}$ and $\Lambda_{low} = 0.6 - 8.2$ W m$^{-1}$ K$^{-1}$, depending on the cycle number and the degree to which it is lithiated.

    $M$ of electrode materials strongly depends on SOC and electrochemical reaction mechanism. We observe reversible lattice stiffening of intercalation materials ($M$ increases with $x$) while conversion and alloying materials show lattice softening ($M$ decreases with $x$) with varying reversibility.

The irreversible $\Lambda$ loss of conversion and alloying electrode materials are closely related to the degree of lithiation (degree of volume expansion required to accommodate lithium). During cycling, conversion and alloying electrode materials can undergo structural degradation and pulverization, creating defects, voids, cracks and SEI within the electrode materials. As these factors are extrinsic to the properties of the electrode material, the observed decrease in $\Lambda$ of the electrode material during cycling is best considered to be a convolution of the material properties and large scale defects which further inhibit the flow of heat [77,78]. The fact that the measured value of $\Lambda$ drops below the minimum thermal conductivity in some cases is strong evidence for defects extrinsic to the electrode material.

Figure 6b shows the relationship between the volume expansion and irreversible $\Lambda$ loss of electrode materials up to five electrochemical cycles. Intercalation electrode materials exhibit small volume expansion and excellent thermal conductivity retention over cycling. Conversion and alloying electrode materials show a large variation of volume expansion depending on their theoretical capacity and cycling condition. Irreversible $\Lambda$ loss increases with the volume expansion of conversion and alloying electrode materials. As thermal conductivity of Sb decreases with increasing number of cycles, Sb exhibits wide Y-ranges.

A simple but effective answer to irreversible thermal conductivity loss of alloying electrode materials would be setting capacity limits below the theoretical capacity to reduce volume expansion and the extrinsic defects. Use of "Single crystal" electrode materials could significantly reduce irreversible thermal conductivity due to their excellent thermal and structural stability over cycling [79].

## 4. Conclusion

We report clear trends in thermal conductivity and elastic modulus of three important classes of electrode materials and provide a comprehensive comparison as a function of their state-of-charge, number of cycles and electrochemical lithium storage mechanism. Thermal conductivity changes range from 1.5 times to a factor of ten during cycling, and we draw the lowest bound of thermal conductivity of electrode materials achieved during cycling. While many of the advances in Li-ion battery technology have been based on new high-power and energy density electrode materials, prior to this report there was little information on the reversible and irreversible thermal conductivity changes during cycling.

This study emphasizes the state-of-charge dependent thermal properties of Li-ion batteries and the nature of volatile thermal conductivity of certain classes of electrode materials. The thermal conductivity of electrode materials is important for engineering design, and the experimental method studied here can be used to characterize changes in the physical properties of electrode materials during cycling. Understanding how the thermal conductivity of electrode materials change during cycling could be used to enhance the window of operation of Li-ion batteries and provide insights into the thermal evolution of Li-ion battery electrode materials. The thermal conductivity contrast of alloying electrode (Sb) is the largest among the reported materials, providing insights into materials with thermally switchable properties.

## 5. Material and methods

*Sample preparation*: Substrates were prepared by plasma-enhanced chemical vapor deposition (PECVD) of α-$SiO_2$ film (~200 nm) on a sapphire substrate (0001). An Al thin-film (~60 nm) was subsequently deposited on the $SiO_2$/sapphire by magnetron sputtering. Then, $Fe_2O_3$, NiO, $V_2O_5$, and $TiO_2$ films were deposited on the Al/$SiO_2$/sapphire substrate using electron beam evaporation (Rocky Mountain Vacuum Tech, Englewood, CO). The base pressure, e-beam voltage and current were set to $10^{-6}$ Torr, 3.4 kV and 20 mA, respectively. The $SiO_2$ on sapphire substrate is optically transparent, and provides a low thermal conductivity barrier to drive most heat flow through the electrode material. The Al film simultaneously acts as the current collector and the optical transducer for TDTR measurement [80]. Al is electrochemically stable above 0.5 V vs. Li/Li$^+$ and chemically compatible with the electrode materials studied here (Supporting Figure S2-3) (Si, Sn and carbon-based anode materials are incompatible with Al due to their low lithiation voltage, and thus not investigated) [81]. To improve crystallinity, the as-deposited electrode samples were annealed with the following conditions: $Fe_2O_3$ in air at 400 °C for 1 h; NiO in Ar at 300 °C for 2 h; $V_2O_5$ in air at 500 °C for 1 h; $TiO_2$ in air at 400 °C for 2 h. The heat treatment improves the initial thermal conductivity of the electrodes (Supporting Information Figure S2). Sb was deposited by thermal evaporation of Sb pellets (1 –3 mm, Kurt J. Lesker, 99.999%) on a Au/Al/SiO$_2$/sapphire substrate. 15-nm Au film was deposited on top of an optically thick Al transducer to improve adherence with Sb as Au can form a thin intermetallic $AuSb_2$ layer [82].

*Electrochemical Measurements*: $Fe_2O_3$, NiO, $V_2O_5$, and $TiO_2$ electrodes were assembled with a liquid cell consisting of 1.0 M $LiClO_4$ in EC (ethylene carbonate)/DMC (dimethyl carbonate) electrolyte (1:1 by vol) and a lithium metal foil (Alfa Aesar, 99.9%) as a counter electrode. The solid-state cell was employed to perform the electrochemical test of Sb with $Li_2S$-$P_2S_5$ solid electrolyte (Sigma Aldrich, 77.5 to 22.5 molar ratio) and an In-Li counter electrode. The solid-state cell was pressed with a vise clamp during measurement. A complete description regarding the design and assembly of the liquid and solid cells can be found in Supporting Figure S3-4. Electrochemical cycling of electrode materials was performed using a portable potentiostat (Bio-Logic SP-200) at room temperature.

*TDTR*: TDTR detects the transient temperature excursion of an Al transducer film in contact with the material of interest from the time-resolved thermoreflectance signal at the modulation frequency $f$. We performed TDTR [83] operated with 785 nm mode-locked Ti-sapphire laser pulses with a full width at half maximum (FWHM) of 10 nm with a repetition frequency of 80 MHz. A polarized beam splitter (PBS) and sharp-edged optical filters with a cut-off wavelength of 785 nm were used to separate spectrally distinct pump and probe beams. The pump beam was modulated by the electro-optical modulator connected to a function generator (SRS DS345) imposing a square wave modulation at $f = 1 - 11$ MHz. The probe beam was modulated by a mechanical chopper operated at 200 Hz. Both pump and probe beams were focused on the Al/SiO$_2$ interphase through the transparent sapphire substrate. We used an optical delay stage to advance the arrival time of the pump pulses with respect to the probe pulses from $t = -20$ ps to $t = 3.6$ ns. We measured the temperature excursion of the transducer film as intensity changes of the probe beam with a RF lock-in amplifier (SRS SR844) synchronized to the function generator at $f$. *In situ* TDTR measurements were performed during galvanostatic cycling of electrode materials. We solved the analytical heat transport model for layered structures with $\Lambda$ of the electrode material as a free parameter. In this model, heat capacity and thickness of each electrode material were interpolated between the lithiated and delithiated states (Supporting Figure S1).

*Figures*

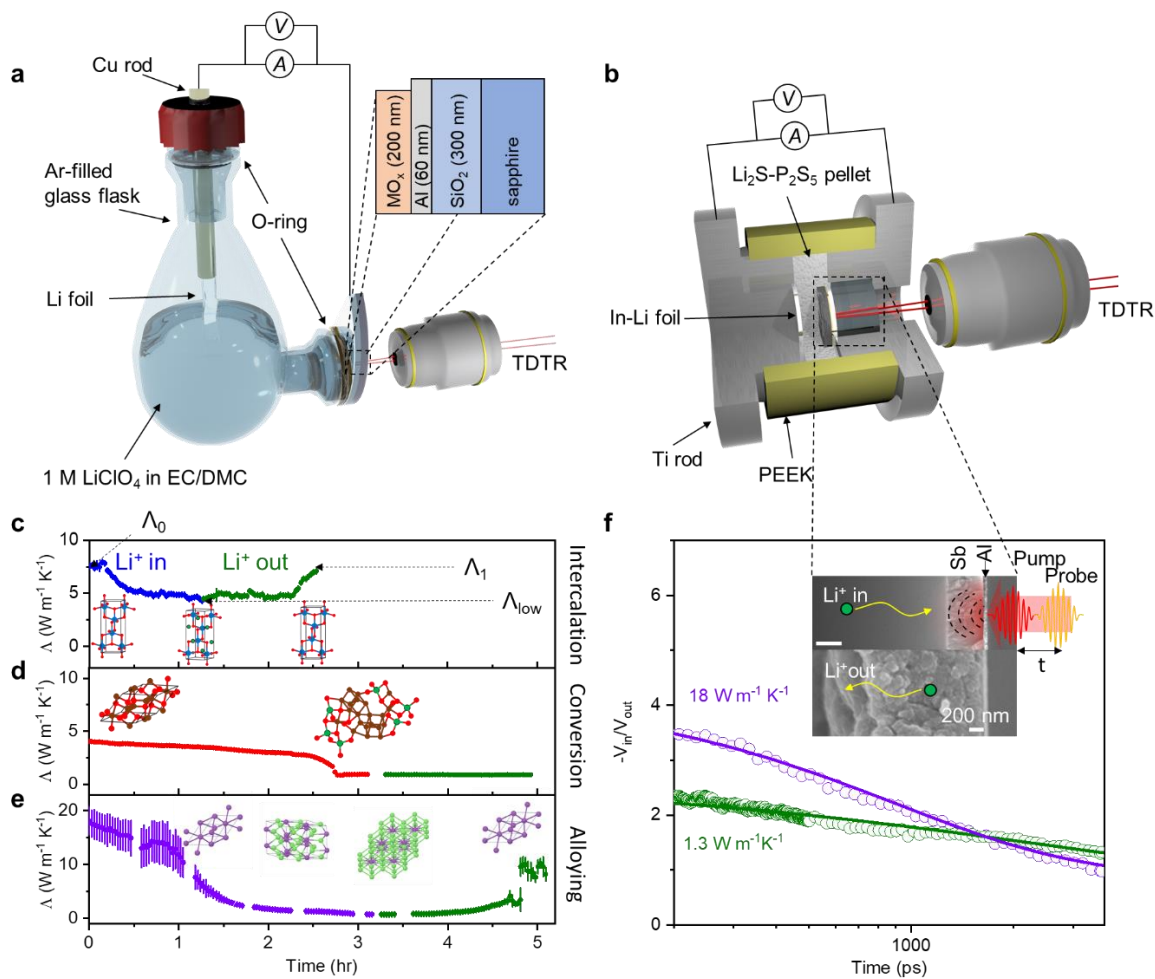

**Figure 1.** *In situ* backside TDTR measurement of intercalation, conversion and alloying electrode materials. (a) Schematic setup of the liquid electrolyte cell with intercalation and conversion metal-oxide ($MO_x$) electrode materials ($TiO_2$, $V_2O_5$, $Fe_2O_3$ and NiO). (b) Schematic setup of the solid-state cell with the alloying electrode material Sb. Features are not drawn to scale. See Supporting information for details. (c)-(e) Thermal conductivity changes of intercalation ($TiO_2$), conversion ($Fe_2O_3$) and alloying (Sb) electrode materials during lithiation and delithiation. (f) Measured and fitted TDTR curves for Sb and fully lithiated Sb. The inset shows cross-sectional scanning electron microscopy (SEM) image of Sb and lithiated Sb electrode

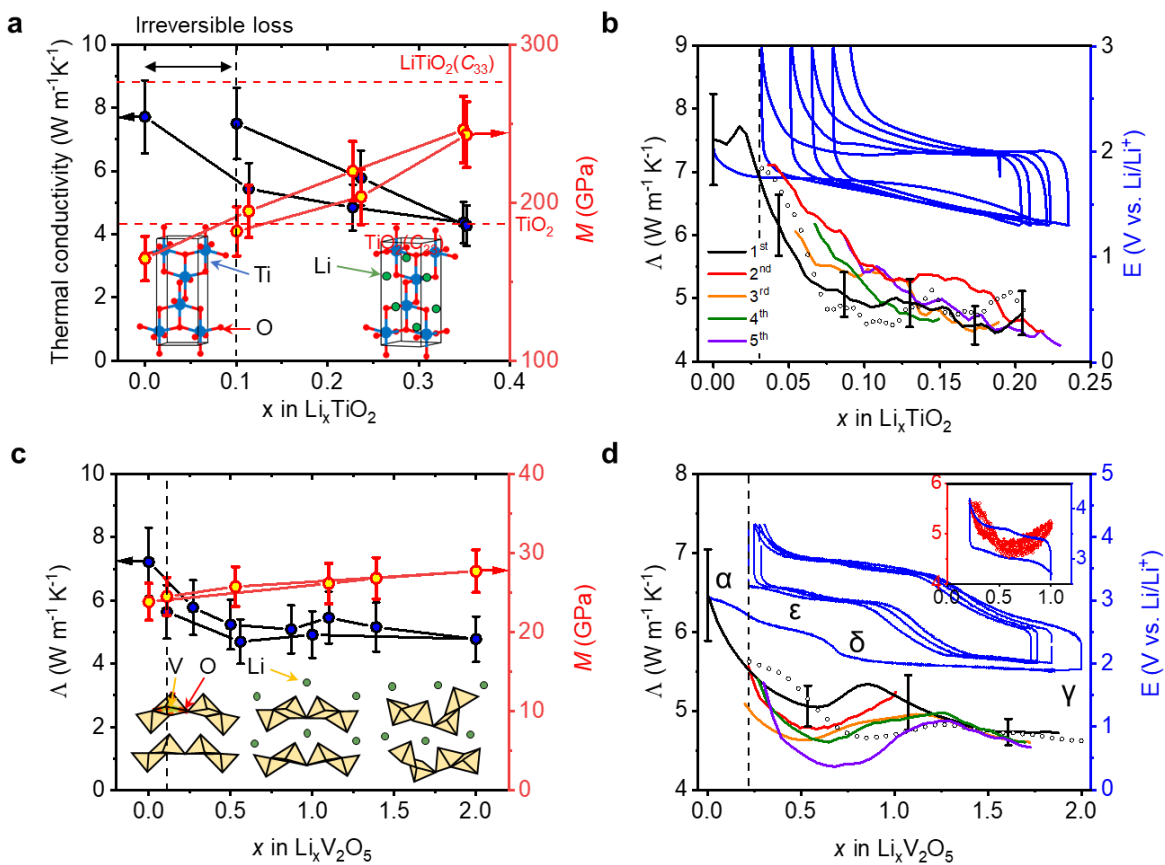

**Figure 2.** $\Lambda$ and $M$ of intercalation $Li_xTiO_2$ and $Li_xV_2O_5$. (a) $\Lambda$ and $M$ of $Li_xTiO_2$ as a function of the lithiation state $x$. The inserted schematic shows the phase transition between the tetragonal $TiO_2$ (left) and orthogonal $Li_xTiO_2$ (right). Li intercalation sites are highlighted with green circles having an occupation probability of $x$. Black dashed lines indicate the irreversible capacity loss at the first discharge. The red dashed line indicates the reference $C_{33}$. (b) $\Lambda$ and potential curves of $Li_xTiO_2$ over five galvanostatic cycles at the current density of 50 mA g$^{-1}$ (3.0 – 1.3 V vs. Li/Li$^+$). (c) $\Lambda$ and $M$ of $Li_xV_2O_5$ as a function of $x$. The inserted graphic illustrates α, δ and γ-$Li_xV_2O_5$ phases. (d) $\Lambda$ and potential curve of $Li_xV_2O_5$ over five galvanostatic cycles at the current density of 100 mA g$^{-1}$. We limit the lithiation of $Li_xV_2O_5$ by setting potential windows differently for each cycle (1st, 1.9 – 4.2 V; 2nd, 2.6 – 4.2 V; and 3rd to 5th, 2.0 – 4.2 V vs. Li/Li+). The inset highlights Λ valley (2nd cycle) over a limited potential window. Open circles show Λ during 1st delithiation.

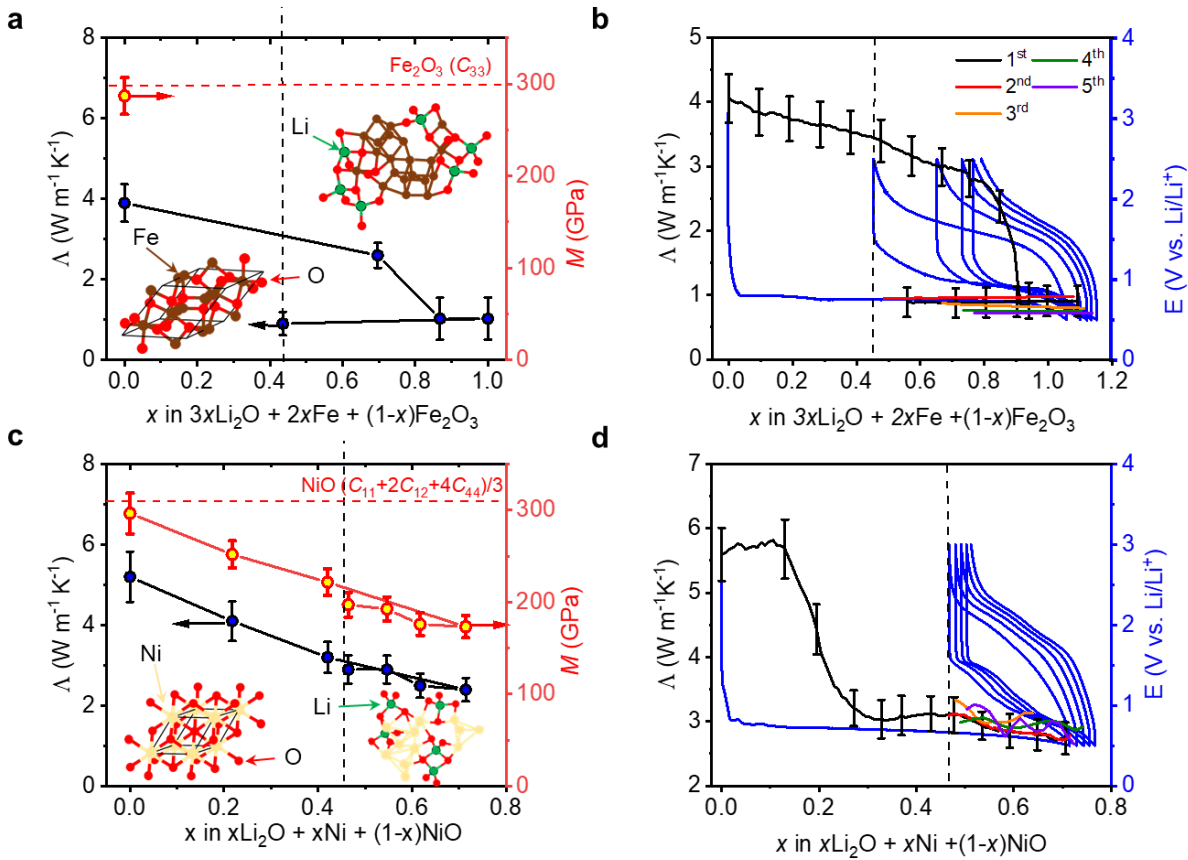

**Figure 3.** Λ and *M* of conversion $Fe_2O_3$ and NiO. (a) Λ and *M* of $Fe_2O_3$ as a function of the lithiation state *x*. *x* of $Li_xTiO_2$ is estimated based on the theoretical electrochemical capacity. Black dashed lines indicate the irreversible capacity loss after the first discharge. The red dashed line indicates the reference $C_{33}$. (b) Λ and cycling curve of $Fe_2O_3$ during five galvanostatic lithiation/delithiation cycles at the current density of 300 mA g$^{-1}$ (2.5 – 0.5 V vs. Li/Li+). (c) Λ and *M* of NiO as a function of *x*. (d) Λ and cycling curve of NiO over five galvanostatic cycles at the current density of 182 mA g$^{-1}$ (3.0 – 0.5 V vs. Li/Li+). The inserted schematics in a and c show the primitive cell of $Fe_2O_3$ and NiO crystals with the nearest atoms and bonds, and their lithiated amorphous structures.

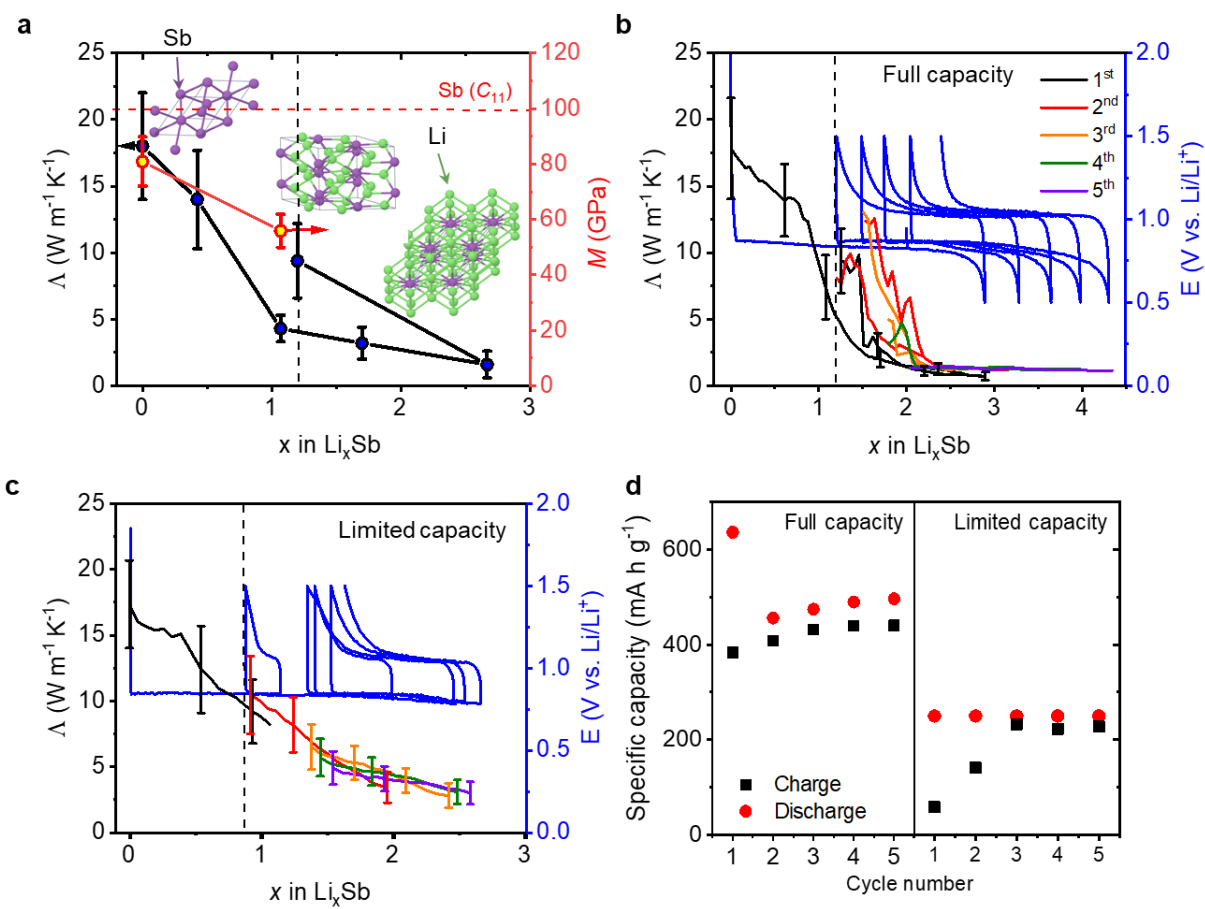

**Figure 4.** Λ and $M$ of alloying Sb. (a) Λ and $M$ of Sb as a function of the lithiation state $x$. The black dashed line indicates the irreversible capacity loss at the first discharge. The red dashed line indicates the reference $C_{11}$ value. The inserted schematics show the crystal structures of Sb, $Li_2Sb$ and $Li_3Sb$. (b) Λ and cycling curve of Sb during five galvanostatic charge/discharge cycles at the current density of 200 mA $g^{-1}$ (1.5 – 0.5 V vs. Li/Li$^+$). (c) Λ and cycling curve of Sb with the limited discharge capacity (250 mA h $g^{-1}$ for each cycle) and the current density of 250 mA $g^{-1}$. Sb is charged to 1.5 V vs Li/Li$^+$ without capacity limit. (d) Electrochemical capacity of Sb with and without the discharge capacity limit.

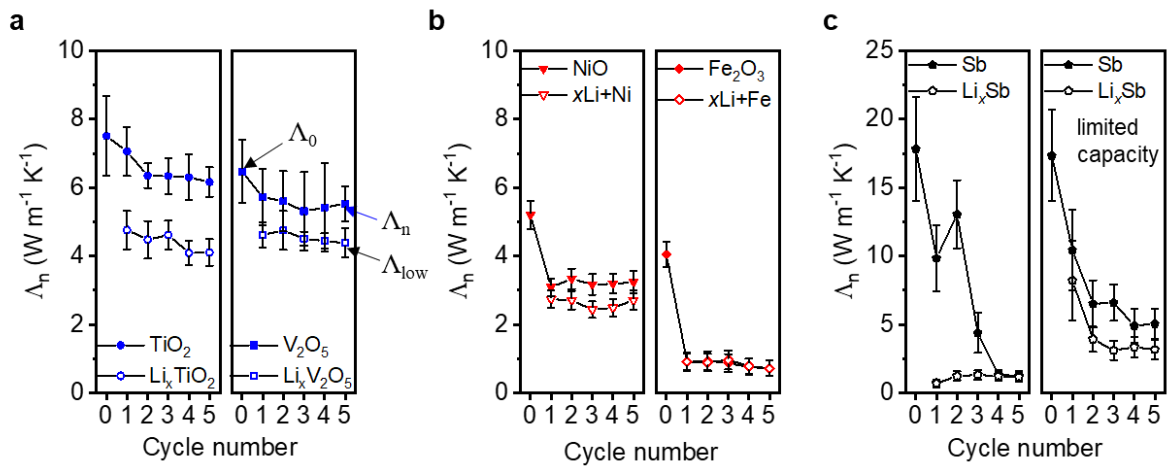

**Figure 5.** Cycle-dependent $\Lambda$ of electrode materials. Thermal conductivity of (a) intercalation, (b) conversion and (c) alloying electrode materials at the initial state ($\Lambda_0$), the lowest state ($\Lambda_{low}$) at each cycle and the delithiated state after completing $n^{th}$ cycle ($\Lambda_n$) plotted.

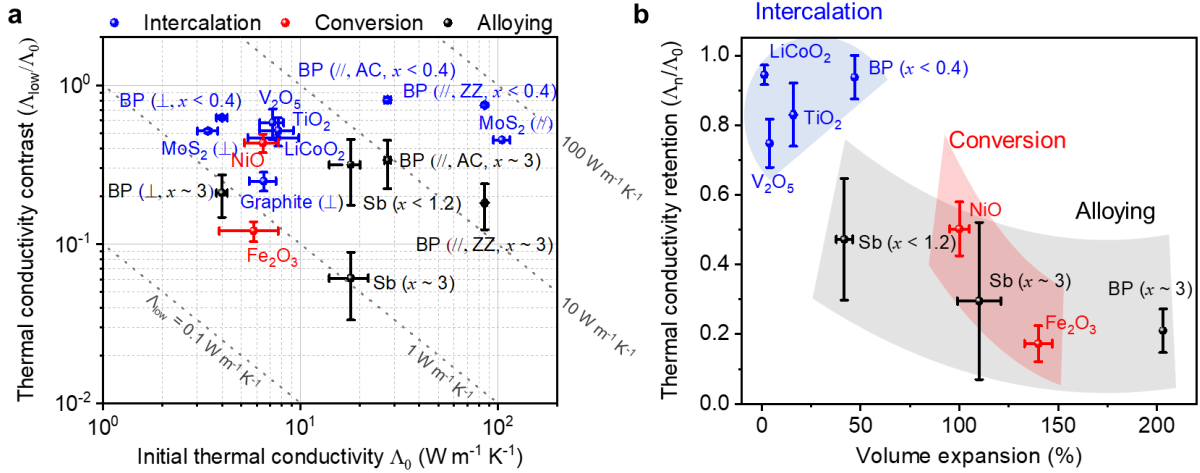

**Figure 6.** SOC dependent $\Lambda$ of electrode materials up to n = 5. (a) $\Lambda$ contrast ($\Lambda_{low}/\Lambda_0$) of Li-ion battery electrode materials. Reference values include through-plane $LiCoO_2$ [13]; in-plane (//) and through-plane ($\perp$) black phosphorus (BP) [16]; and zig-zag in-plane (ZZ), armchair in-plane (AC) and through-plane ($\perp$) $MoS_2$ [14]; and through-plane ($\perp$) graphite [44,45]. X-error bars show sample variations and uncertainties of $\Lambda_0$. Y-range shows the uncertainty and $\Lambda_{low}$ with varying cycle numbers. The diagonal lines are labeled with different $\Lambda_{low}$ to estimate the lower bound of the $\Lambda$ of electrode materials during cycling. (b) Volume expansion and $\Lambda$ retention ($\Lambda_n/\Lambda_0$) of electrode materials over cycling with n = 1 ~ 5. The volume expansion of $LiCoO_2$ is estimated from *in situ* XRD [84]. The volume expansion of BP is calculated from a first-principle study [76]. X-error bar shows experimental uncertainty of volume expansion. Y-error bars show the range of $\Lambda_n$ observed over a number of cycles.

**Acknowledgements**

Samples were prepared at the Research Institute of Advanced Materials at Seoul National University and the Materials Research Laboratory at the University of Illinois Urbana–Champaign. Materials, thermal and electrochemical characterization were performed at the Materials Research Laboratory at the University of Illinois Urbana–Champaign. This work was supported by the National Science Foundation Engineering Research Center for Power Optimization of Electro-Thermal Systems, with Cooperative Agreement EEC-1449548 and the US Army CERL W9132T-19-2-0008. S.K. appreciates the Kwanjeong Educational Foundation scholarship.

**Conflict of interests**

The authors declare no competing interests.

**Author contributions** J.S., S.K., D.G.C. and P.V.B. conceived the idea. J.S. and S.K. designed the liquid and solid electrochemical cells. J.S. prepared the substrates. H.P., H.J. and S.K. deposited the electrode materials. S.K. conducted thermal annealing of the samples and assembled the cells. J.S. conducted TDTR, picosecond acoustics and RBS measurements. S.K. performed SEM, XRD and electrochemical experiments. J.S., S.K., D.G.C. and P.V.B. analyzed the data. J.S., S.K., D.G.C. and P.V.B. wrote the manuscript. All authors read and edited the manuscript. D.G.C. and P.V.B supervised the research.

# Supporting Information

**Thermal conductivity of intercalation, conversion, and alloying lithium-ion battery electrode materials as function of their state of charge**

*Jungwoo Shin, Sanghyeon Kim, Hoonkee Park, Ho Won Jang, David G. Cahill\*, Paul V. Braun\**

**Time-domain thermoreflectance (TDTR)**

*In situ TDTR measurement*: Thermal conductivity ($\Lambda$) change of each electrode was measured by continuous recording of the in-phase and out-of-phase voltages ($V_{in}$ and $V_{out}$) of an RF lock-in amplifier under galvanostatic cycling where $-V_{in}/V_{out}$ is related to the phase angle $\Phi = \arctan(V_{out}/V_{in})$. With a bidirectional heat transfer model[1] $\Lambda$ was solved where the heat capacity per unit volume $C$ and the thickness $d$ of the electrode materials were linearly interpolated between the lithiated and delithiated states at the lithiated state $x$. During *in situ* TDTR measurement, the current and electrochemical potential of electrode materials were simultaneously controlled and measured by a portable potentiostat (SP-200, Bio-Logic).

We calculate the sensitivity $S_X$ as the ratio of the change in a thermal quantity $X$ to a change in the TDTR signal ($-V_{in}/V_{out}$). $X$ can be any thermal parameters such as $\Lambda$, $C$ or $d$ of each layer.

$$S_X = \frac{d\ln\left(-\frac{V_{in}}{V_{out}}\right)}{d\ln X} \qquad (6)$$

As $S_X$ is sensitive to the modulation frequency $f$, beam spot size $w$, $d$ and $C$ of each layer, we optimize the experimental parameters in which $S_\Lambda$ of the electrode materials maximizes relative to others. We set the thickness of materials investigated to less than 300 nm to enable facile lithium diffusion at a moderate C-rate (> 0.2 C). With the exception of pristine Sb, we ensure the material thickness in the delithiated state is at least the thermal penetration depth $[\Lambda/(\pi f C)]^{1/2}$, which is calculated as 200 – 300 nm for $TiO_2$, $V_2O_5$, $Fe_2O_3$ and NiO at the highest modulation frequency $f$ = 11 MHz. By keeping the electrode thicker than the thermal penetration depth, the thermal conductivity of the electrolyte becomes unimportant. The thickness of pristine Sb (200 – 300 nm) was smaller than the thermal penetration depth (650 nm) at the initial state due to the high $\Lambda$ ~18 W $m^{-1}$ $K^{-1}$. $\Lambda$ of Sb at the initial state is independently confirmed by front-side TDTR measurement. During lithiation, most materials expand and their $\Lambda$ decreases, so during cycling we do not move into a regime where the thermal penetration depth exceeds the material thickness. Therefore, the liquid and solid electrolyte do not affect TDTR signal during cycling.

Unlike intercalation electrode materials with changes in $C$ and $d$ less than 10% during cycling, conversion and alloying electrodes show non-negligible changes in $C$ and $d$ during cycling. These changes must be considered in the modeling, as they impose an obvious shift in $S_\Lambda$ with respect to $x$. All $C$ values at the lithiated and delithiated electrodes were measured by front-side TDTR with varying $f$ except Sb. Due to the irregular surface thickness, we were not able to confirm $C$ at the lithiated state of Sb. Instead, we used a calculated heat capacity of $Li_3Sb$ $C$ = 2.36 J $cm^{-3}$ $K^{-1}$ where both Sb and Li match well with the classical heat capacity (Note, the Debye temperature of $Li_3Sb$ is = 171 K). Figure S1 shows sensitivity curves and measured and fitted TDTR curves for $Fe_2O_3$, NiO and Sb during cycling.

Note that $\Lambda$ of lithiated Sb could be underestimated as Sb shows the largest heat capacity transition $\Delta C$ = 70% during cycling. Defect, void and inhomogeneous phase transition could result in a discrepancy between actual $C$ and the interpolated $C(x)$, which will underestimate $\Lambda$ to the same extent ($S_\Lambda \sim S_C$, Figure S1i). The solid electrolyte of Sb is considered as a thermally thick layer where the thermal effusivity $\varepsilon = (\Lambda C)^{-1/2}$ of the solid electrolyte was fixed to 1500 W s$^{1/2}$ m$^{-2}$ K$^{-1}$.

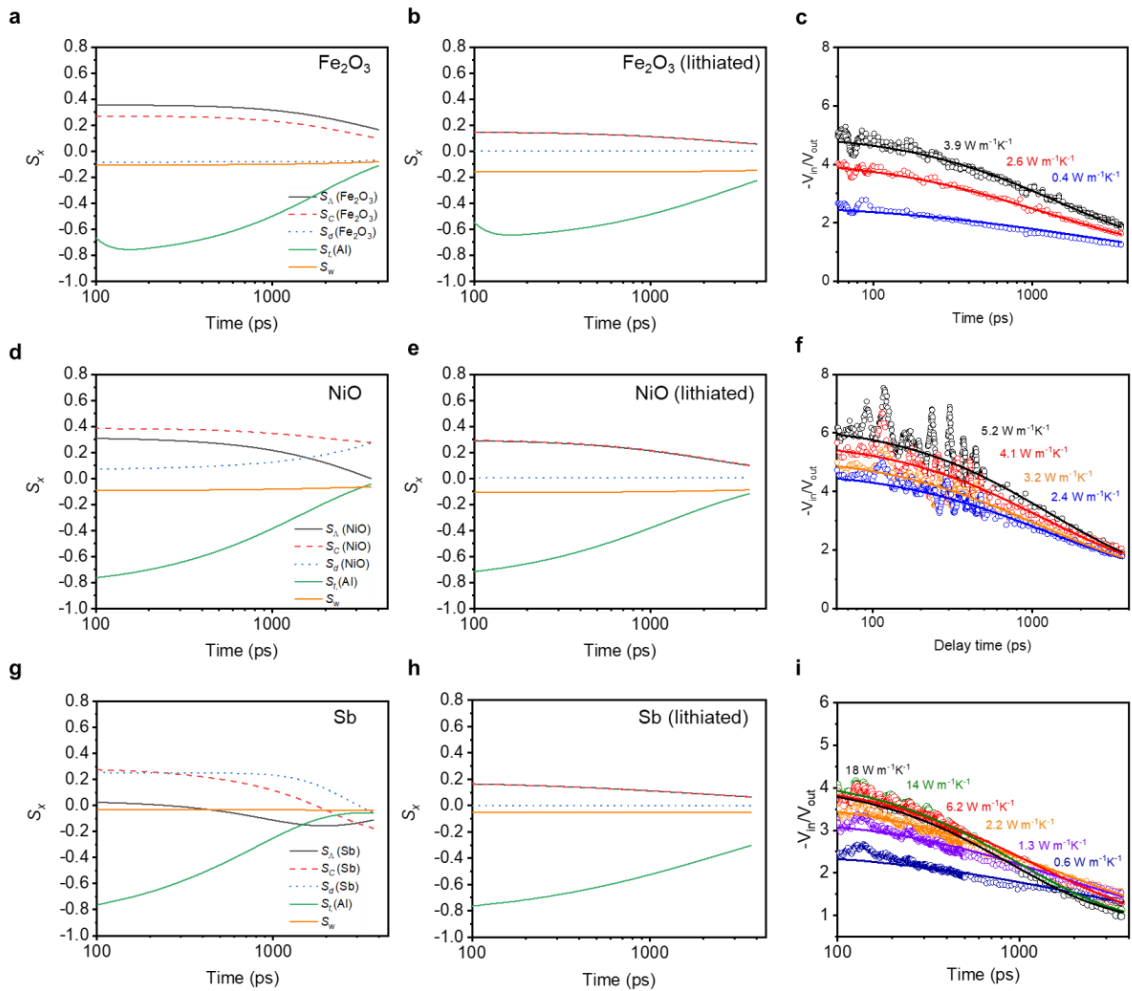

**Figure S1.** Sensitivity analysis of electrode materials. (a)-(b) Calculated sensitivity curves of Fe$_2$O$_3$ at the initial and lithiated state. (c) Measured and fitted TDTR curves for Fe$_2$O$_3$ during cycling. (d)-(e) Calculated sensitivity curves of NiO at the initial and lithiated state. (f) Measured and fitted TDTR curves for NiO during cycling. (g)-(h) Calculated sensitivity curves of Sb at the initial and lithiated state. **i** Measured and fitted TDTR curves for Sb during cycling.

*Ex situ TDTR measurement*: We carried out *ex situ* TDTR measurements to confirm the remarkably large $\Lambda$ decrease of $Fe_2O_3$ and Sb after the first lithiation process. To maximize the initial $\Lambda$, we deposited $Fe_2O_3$ film on Au (60 nm)/$SiO_2$ substrates with varying annealing temperatures up to 500 °C in air for 1 h. After the heat treatment, we fully discharged $Fe_2O_3$ in an Ar-filled glove box and deposited Al film on the sample for *ex situ* TDTR measurement. Prior to Al deposition, samples were sealed in the Ar-filled container and immediately brought to the Al deposition chamber to minimize exposure to air and moisture. Regardless of the initial $\Lambda$ of $Fe_2O_3$ (1.7 – 7.6 W m$^{-1}$ K$^{-1}$, depending on the heat treatment condition), we observed a similar $\Lambda$ of $Fe_2O_3$ after the first lithiation (0.3 – 1.1 W m$^{-1}$ K$^{-1}$), suggesting that $\Lambda$ at the lithiated state is independent of the initial $\Lambda$. Similarly, we repeated TDTR measurement of multiple Sb samples with varying deposition rates ($\Lambda$ = 2.6 – 18 W m$^{-1}$ K$^{-1}$) and confirmed thermal conductivity reduction is universal. Note that the reason for the $\Lambda$ discrepancy of $Fe_2O_3$ between *ex situ* TDTR and *in situ* TDTR measurements (7.6 vs 3.9 W m$^{-1}$K$^{-1}$) is the difference of annealing temperature (400 vs. 500 °C).[2]

*Effect of optical artifacts*: The thermoreflectance signal is subject to optical artifacts due to non-negligible thermo-optic constant (d*n*/dT) of materials in the beam path and beyond. In this study, the electrolytes exhibit large d*n*/dT, which can alter the amplitude of the thermoreflectance signal when the transducer is not opaque. Since our TDTR model assumes that the reflected probe beam only carries the information on the temperature change of the transducer layer, an oscillation of the beam reflected from the electrolyte or back substrate synchronized with *f* would end up with overestimation or underestimation of $\Lambda$ of the electrode materials.

Since $d$ of the transducer film for the backside TDTR measurement is inversely proportional to $S_\Lambda$ of electrode materials, there is an optimum range of the transducer thickness to achieve high enough sensitivity while securing optical opacity. Here, we chose $d = 60$–$80$ nm as an optimum thickness of the Al transducer layer to achieve high enough $S_\Lambda$ for the backside TDTR. In this range, Au and Pt transducers are partially transparent and their thermoreflectance ($dR/dT$) values are about an order of magnitude smaller than Al at $\lambda = 785$ nm.[3] Therefore, Au and Pt are subject to a significant optical artifact (~ 20% of the total TDTR signal) based on the fraction of the beam reflected from the electrolyte with a high $dn/dT > 10^{-4}$ K$^{-1}$. Therefore, we chose Al as a transducer film and a current collector for working electrode materials for the backside TDTR measurement.

To estimate the optical artifact induced by the $SiO_2$ film on the substrate we performed an optical transfer-matrix calculation with Al/$SiO_2$/sapphire to calculate the effect of thermo-optic constant of the substrate. Given the thermo-optic constant of $SiO_2$ ($dn/dT = 8.6 \times 10^{-6}$ K$^{-1}$ at 785 nm),[4] the total thermoreflectance was shifted by up to 0.7 % compared with the ideal substrate ($dn/dT = 0$) considering the transient temperature rise during the measurement. From the sensitivity analysis, we estimate the effect of optical artifacts on $\Lambda$ uncertainty would not exceed 2 %.

*Electrochemical stability of substrate*: As Al forms a binary alloy with Li by an electrochemical reaction below 0.5 V vs Li/Li$^+$, we chose electrode materials whose working potential window is higher than 0.5 vs Li/Li$^+$.[5] This potential range works with all cathode materials and a limited number of anode materials for lithium-ion (Li-ion) batteries. Among anode materials, 3-$d$ metal oxides, Sn and Sb-based anode materials are stable in this range, but carbon and silicon-based anode materials exhibiting discharge potential plateaus below 0.5 V vs Li/Li$^+$ are incompatible with Al film.

Here, we tested the voltage window of the Al/SiO$_2$/sapphire substrate shows cyclic voltammetry (CV) curves with varying cut-off voltages from 0.5 to 0 V vs Li/Li$^+$. Figure S2 shows that the Al film is stable above 0.2 V vs Li/Li$^+$, which is slightly lower than the standard potential of Li-Al reaction attributable to the presence of a thin layer of native oxide.

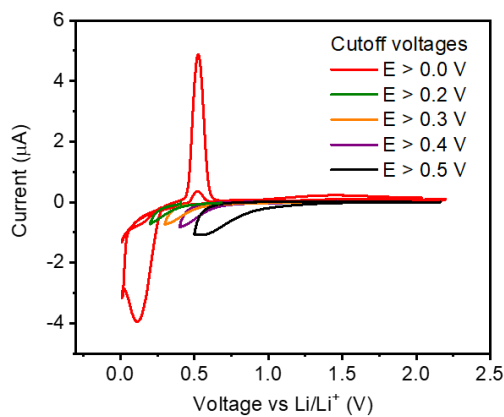

**Figure S2.** Electrochemical stability of Al film. Whereas CV curves show no specific redox peak above 0.2 V vs Li/Li$^+$, we found electrochemical redox peaks of Al film below 0.2 V vs Li/Li$^+$, indicating that Al reacts with Li$^+$ ions below 0.2 V vs Li/Li$^+$. A Li-metal foil was used as a counter electrode and 1 M LiClO$_4$ dissolved a mixture of ethylene carbonate (EC) and dimethyl carbonate (DMC) (1:1 by volume) was used as a liquid electrolyte.

*Thermal stability of substrate*: We annealed samples to achieve highest the crystallinity and thermal conductivity at the initial state. As backside TDTR measurements require the electrode materials to be deposited on the Al/SiO$_2$/sapphire substrate, we first ensured that the heat treatment does not alter the optical property of Al film. Figure S3a shows measured and fitted TDTR data of Al/SiO$_2$/sapphire substrates before and after the heat treatment for 2 h at 400 °C in air. We attributed a slight change in the TDTR signal to the growth of the native oxide on the Al substrate, from 4.2 nm to 7.0 nm, confirmed by ellipsometry (Gaertner L116C) and TDTR fitting. We further confirmed the thermal stability of Al film by varying annealing temperature. Where Al film exposed to air was damaged at above 400 °C, we confirmed that Al film covered with $V_2O_5$ survived at an elevated temperature up to 500 °C for 1 h.

Figure S3b-c shows TDTR data of the $V_2O_5$ and $Fe_2O_3$ before and after the heat treatment. We measured that the optical reflectivity of as-prepared and heat-treated $V_2O_5$/Al/$SiO_2$/sapphire substrate remained unchanged but the thermal conductivity of $V_2O_5$ increased from 0.9 W m$^{-1}$ K$^{-1}$ to 8.4 W m$^{-1}$ K$^{-1}$ due to the improved crystallinity and the reduction of defects.[6] Similarly, $\Lambda$ of $Fe_2O_3$ increased from 1.7 W m$^{-1}$ K$^{-1}$ to 3.9 W m$^{-1}$ K$^{-1}$ (400 °C) and 7.6 W m$^{-1}$ K$^{-1}$ (500 °C). $\Lambda$ of NiO increased from 2.1 W m$^{-1}$ K$^{-1}$ to 5.2 W m$^{-1}$ K$^{-1}$ (300 °C).

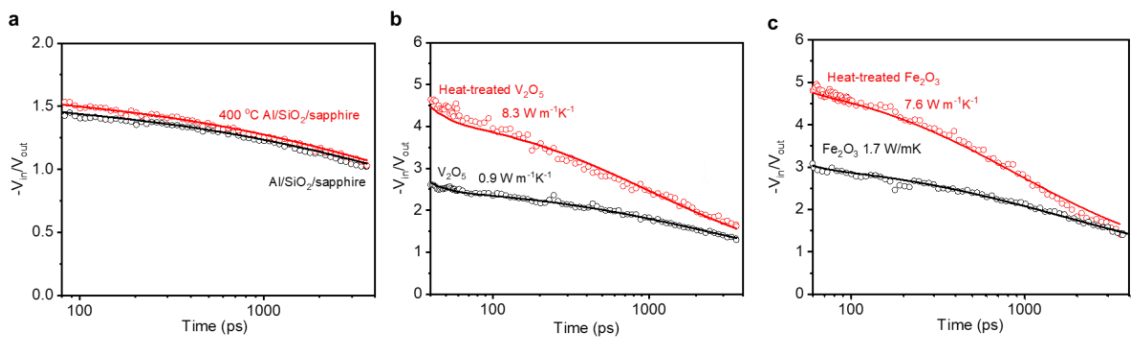

**Figure S3.** The effect of heat treatment. (a) Measured (open circles) and fitted (solid lines) TDTR curves of Al/$SiO_2$/sapphire before and after the heat treatment at 400 °C in air. (b) Measured (open circles) and fitted (solid lines) TDTR curves of $V_2O_5$ before and after the heat treatment at 500 °C in air. (c) Measured (open circles) and fitted (solid lines) TDTR curves of $Fe_2O_3$ sample before and after the heat treatment at 500 °C in air.

*Steady-state temperature rise*: We used Al/$SiO_2$/sapphire substrates for backside TDTR measurements of electrode materials. The thin $SiO_2$ film (200 – 300 nm) was set to thicker than the heat penetration depth, thereby increasing $S_\Lambda$ of the electrode materials.[2, 7] The thickness of the SiO$_2$ layer plays a dominant role in the steady-state temperature rise $\Delta T_{ss}$ due to the accumulation of heating pulses. In this paper, we used 5× and 10× objective lenses to focus the pump and probe beam onto the Al film with 1/e$^2$ intensity radii of 10 and 5 μm, respectively. The intensities of the pump and probe beam were set to 5–10 mW and 3–5 mW, respectively. The corresponding temperature rise by a single pump pulse $\Delta T_{pp}$ and $\Delta T_{ss}$ are calculated as ~ 1.5 K and ~ 8 K, respectively.

Sapphire was used as a high-thermal-conductivity heat sink. We calculate that the contribution of steady-state temperature rise added by the SiO$_2$ layer is 50 % by comparing $\Delta T_{ss}$ with and without the sandwiched SiO$_2$ layer. Replacing thin SiO$_2$ with a thick SiO$_2$ layer would result in a dramatic increase in $\Delta T_{ss}$ > 30 K, which is not appropriate for the accurate $\Lambda$ analysis.

**Design of electrochemical cell**

*Liquid cell*: Liquid cells were prepared by a 25-mL round bottom flask modified with a 1-cm diameter opening with an O-ring joint on the side (Figure S4). The opening was clamped with a Viton O-ring and an electrode/Al/SiO$_2$/sapphire substrate. Then, the liquid cell was filled with a 1 M LiClO$_4$ dissolved in ethylene carbonate (EC) and dimethyl carbonate (DMC) (1:1 by volume) electrolyte and assembled with a Li metal foil counter electrode (Alfa Aesar, 99.9%) for the two-electrode setup in an Ar-filled glove box. A Cu rod modified with an alligator clip was used to hold the Li foil inside the flask, sealed with a modified 24/40 thermometer adapter and a keck clamp.

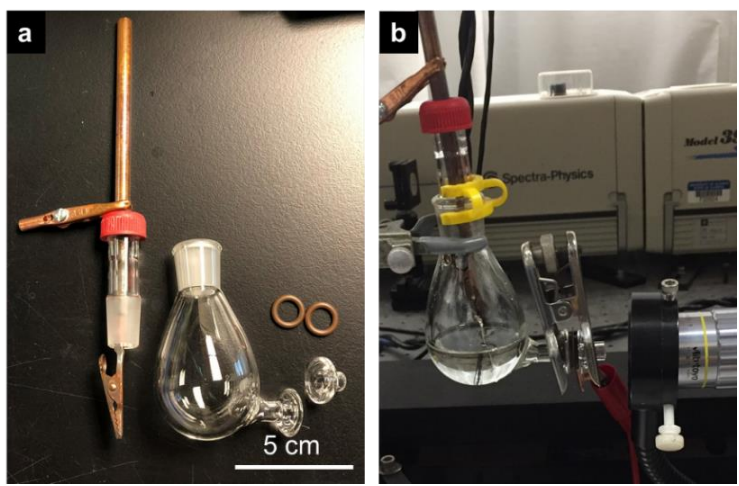

**Figure S4.** Liquid electrochemical cell assembly. (a) A photograph of the components for liquid electrolyte cell assembly including a modified 25-mL round flask, an O-ring joint, O-rings and a modified 24/40 thermometer adapter with a Cu rod and an alligator clip. (b) A photograph of the liquid electrolyte cell during *in situ* backside TDTR measurement.

*Solid-state cell*: We prepared the solid-state cell consisting of polyaryletheretherketone (PEEK) cell die with two machined Ti rods that serve as current collectors for both working and counter electrodes (Figure S5). One of the Ti rods was modified with a stepped hole (1 - 2 mm) to load a sample and to provide the laser beam pathway. We prepared a $Li_2S$-$P_2S_5$ solid electrolyte by high-energy ball milling for 10 h (8000M Mixer/Mill, SPEX SamplePrep) with a 77.5 to 22.5 molar ratio $Li_2S$ (Sigma, 99.98%) and $P_2S_5$ (Sigma, 99%) =in a stainless-steel vial under Ar atmosphere.

A complete procedure of the solid-state cell assembly is as follows. In the Ar-filled glove box, Sb/Au/Al/$SiO_2$/sapphire substrate (5 mm diameter) was inserted into the stepped hole in the modified Ti rod. Then, we uniformly spread 100-mg 77.5$Li_2S$-22.5$P_2S_5$ solid electrolytes on the Ti rod mounted with the PEEK cell die. Then, we pressed the electrolyte (30 MPa) to form a 13 mm diameter pellet. Finally, the indium-lithium (In-Li) counter electrode was attached to the other side of the solid electrolyte pellet and pressed at 30 MPa with the other Ti rod. The solid-state cell was sealed with a 3M tape before taken out of the glove box. During the TDTR measurement, a steel vise was used to keep the pressure of the solid-state cell (Figure S5c).

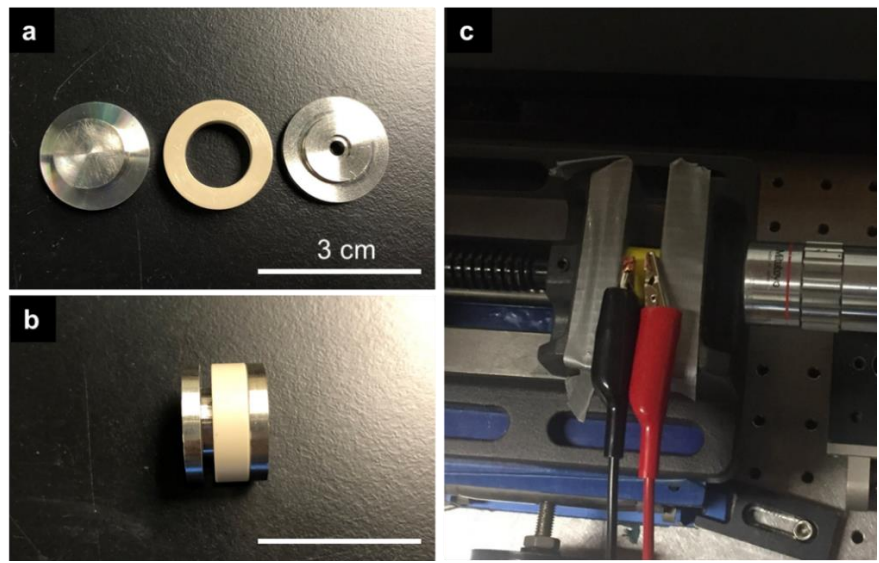

**Figure S5.** Solid-state cell assembly. (a) A photograph of solid-state cell components including two machined Ti rods and a PEEK cell die. Note that one Ti rod is modified with a stepped hole to load a 5-mm diameter sample. A 2-mm opening provides the laser beam pathway. The transparent side of an Sb/Au/Al/SiO$_2$/sapphire substrate works as an optical window where the pump and probe beams were focused onto the Al/SiO$_2$ interphase through the stepped hole of the modified Ti rod. (b) A photograph of an assembled solid-state cell before pressurization. (c) A photograph of a solid-state cell during *in situ* backside TDTR measurement. The solid-state cell was kept pressurized with a steel vise to prevent the delamination of the Sb film.

**Structural transition of electrode materials**

*Volume expansion*: We measured volume changes of the five electrode materials between the lithiated and delithiated states using cross-sectional scanning electron microscopy (SEM). Figure S6 shows cross-sectional SEM images of as-prepared (after the heat treatment), lithiated and delithated electrode materials deposited on Al/$SiO_2$/sapphire substrates. We compared the thicknesses before and after the first lithiation reaction. $TiO_2$ and $V_2O_5$ exhibit theoretical volume expansions of less than 6% upon lithium intercalation.[8] The experimental volume expansion of intercalation $TiO_2$ (anatase) was calculated as 16%, slightly larger than the theoretical volume. The additional volume expansion of $TiO_2$ could be attributed to the formation of SEI (solid electrolyte interface). After the first delithiation process, $TiO_2$ showed a small irreversible volume expansion of 3%, which is the estimated comparing the thickness of the initial and delithiated states. $V_2O_5$ showed a negligible volume expansion of < 1 % during charge/discharge cycles. In comparison, conversion $Fe_2O_3$ showed a volume expansion of 140%. The irreversible volume expansion ratio was measured as 16%. NiO showed a slightly lower volume expansion of 100% with a slightly higher irreversible volume expansion of 22%. The alloying Sb showed the most dramatic volume expansion accompanied by severe structural degradation, which leads to high uncertainty in the volume expansion estimation. Note that cycling in a liquid electrolyte cell resulted in immediate delamination of Sb film due to the dramatic volume expansion of the film. Even with the solid-state cell, Sb film was highly deformed after cycling (after 10 cycles, Figure S6b). This surface deformation is a common phenomenon for alloying electrode materials with a large volume expansion.[9]

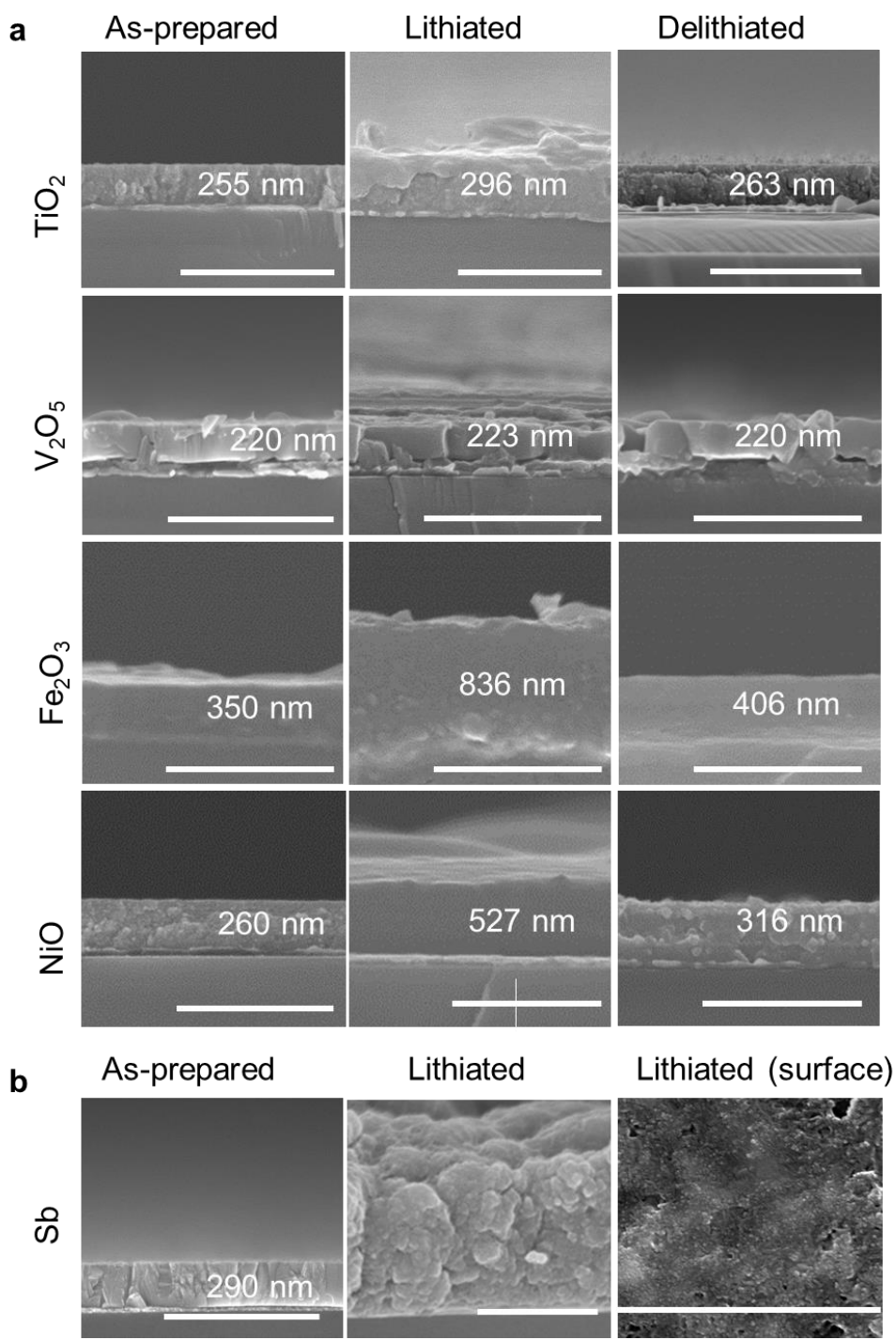

**Figure S6.** Volume expansion of electrode materials. (a) Cross-sectional SEM images of as-prepared, lithiated and delithiated conversion and intercalation electrode materials on Al/SiO$_2$/sapphire substrates. (b) Cross-sectional and surface SEM images of as-prepared and lithiated (10$^{th}$ cycle) Sb/Au/Al/SiO$_2$/sapphire. The scale bar is 1 μm.

*Phase transitions*: Metal to oxygen stoichiometry and crystalline phase of electrode materials were characterized by Rutherford backscattering spectrometry (RBS) and grazing incidence X-ray diffraction (GI-XRD). First, we performed RBS using a high-energy He$^+$ ion beam (0.5-2.0 MeV) emitting from a high-energy ion-beam accelerator (3SDH Pelletron, NEC). GI-XRD was performed using Panalytical Phillips X'pert with a Cu-Kα radiation (λ = 0.154 nm). Figure S7 shows measured and fitted RBS curves for the intercalation and conversion electrode materials deposited on Al/SiO$_2$/sapphire substrates. Figures S7a-b shows RBS data of TiO$_2$ and V$_2$O$_5$ on Al/SiO$_2$/sapphire substrates where the fitted metal-to-oxygen ratios are Ti:O = 31.4:68.6 and V:O = 28.7:71.3, close to the theoretical values of Ti:O (Ti:O = 33.3:66.7) and V$_2$O$_5$ (V:O = 28.6:71.4). On the other hand, the calculated atomic ratio of Fe$_2$O$_3$ electrode is Fe:O = 41.4:58.5 (Fe$_{2.2}$O$_3$), suggesting the presence of Fe-rich phases such as Fe$_3$O$_4$ (Figure S6c). RBS signal fluctuation of Ni peak of NiO sample (Figure S7d) also suggested an oxygen gradient with varying atomic ratios of Ni:O = 55.2:44.8 (bottom) 65:35 (middle) and 50:50 (top), indicating incomplete oxidation of NiO in bulk during the heat treatment.

Figure S8 shows GI-XRD data of as-prepared, lithiated and delithiated five electrode materials. First, we measured each sample before the cell assembly. After *in situ* TDTR measurements we disassemble the cell and recollect the samples in the Ar-filled glove box. To be specific, samples at the delithiated state were cut into two pieces and one of each sample was lithiated in the glove box with the identical current density used in the *in situ* TDTR measurements. Then, the lithiated and delithiated samples were sealed with Kapton tapes in the glove box and immediately brought for the XRD measurements.

Intercalation electrode materials exhibited crystalline structures during cycling. Figure S8a shows XRD data of as-prepared, lithiated and delithiated $TiO_2$. Among $TiO_2$ polymorphs, we confirmed anatase $TiO_2$ (α phase). Upon lithiation, we observed the (020) diffraction peak shift from $2\theta$ = 48.33 (b = 3.78 Å) to 44.68° (4.054 Å), associated with the phase transition between α-$TiO_2$ (b = 3.792 Å) and β-$Li_xTiO_2$ (b = 4.084 Å), respectively.[10] Figure S8b shows XRD data of $V_2O_5$. We confirmed that α-$Li_xV_2O_5$ at the initial state was transformed to a binary mixture of δ-$Li_xV_2O_5$ (x < 1) and γ-$Li_xV_2O_5$ (1 < x ≤ 2) phases after the lithiation, and then returned to the initial α-$Li_xV_2O_5$ phase after the delithiation process.[11]

On the other hand, conversion electrodes do not show reversible crystal transition. We observe diffraction peaks corresponding to (c) $Fe_2O_3$ and $Fe_3O_4$ phases and (d) NiO and Ni phases, consistent with the off-stoichiometries observed in RBS measurements. After the first discharge, we observe a loss of crystal structure of $Fe_2O_3$ and NiO indicating irreversible loss of crystallinity where the amorphous metal and $Li_2O$ do not give rise of diffraction peaks. Delithiation does not recover the crystalline structures.

Lastly, Figure S8e shows XRD data of Sb at the initial state and the lithiated state. XRD data suggest that Sb turned into $Li_3Sb$ after the lithiation process. A broad peak at $2\theta$ = 30° indicates the presence of the remaining $Li_2Sb$ phase. The diffraction peaks of $Li_2S$ originated from a trace of $Li_2S:P_2S_5$ solid-electrolyte on the surface of the recovered electrode where $Li_2S:P_2S_5$ solid-electrolyte can undergo phase decomposition into $Li_2S$ during cycling.[12] Note that Sb film strongly adheres to the solid electrolyte at the delithiated state, which makes it difficult to detect the diffracted signal of Sb film after cycling.

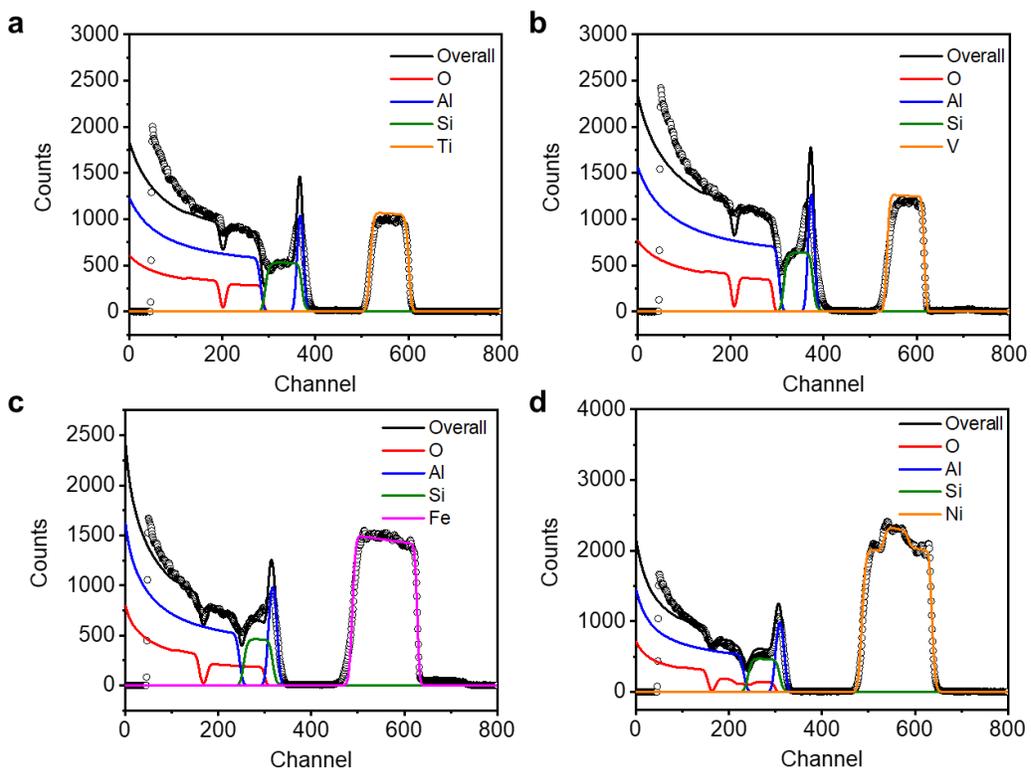

**Figure S7.** Measured and fitted RBS data of intercalation and conversion electrode materials. (a)-(b) Intercalation $TiO_2$ and $V_2O_5$ on Al/SiO$_2$/sapphire substrates. (c)-(d) Conversion $Fe_2O_3$ and NiO on Al/SiO$_2$/sapphire substrates.

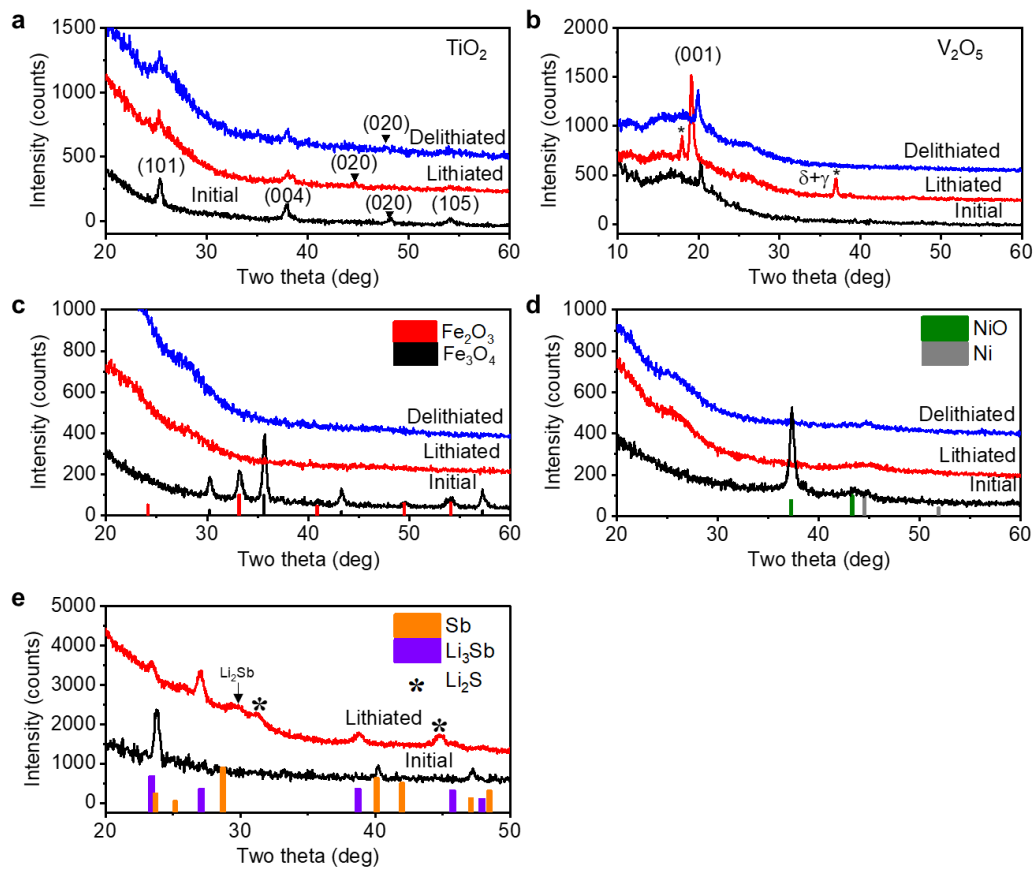

**Figure S8.** XRD data of five electrode materials before and after lithiation/delithiation reaction. (a)-(b) Intercalating $TiO_2$ and $V_2O_5$. (c)-(d) Conversion $Fe_2O_3$ and NiO. (e) Alloying Sb.

**Elastic modulus of electrode materials**
*Speed of sound*s: We measured the speed of sounds of electrode materials with varying lithium contents $x$ using picosecond acoustic oscillation. The picosecond acoustic signals are generated from the reflection of acoustic waves between (i) Al/SiO$_2$ and SiO$_2$/sapphire interphases and (ii) Al/electrode and electrode/electrolyte interphases. Note that severe deformation of electrode materials or contact with materials having similar acoustic impedance ($Z$) will result in no detectable echoes. The acoustic signal of the lithiated Fe$_2$O$_3$ was not detected due probably to the pulverization and large structural deformation.

Figure S9 shows the measured in-phase lock-in amplifier voltage ($V_{in}$) of (a) TiO$_2$, (b) V$_2$O$_5$ and (c) NiO films on Al/SiO$_2$/sapphire substrate in the liquid electrolyte cell and (d) Sb films on Au/Al/SiO$_2$/sapphire substrate in the solid-state cell during *in situ* TDTR measurements. Peaks and valleys marked with a dashed black line correspond to the echoes from SiO$_2$ (type i), which remains constant during cycling. The second acoustic signals (type ii) marked with black triangles vary with electrode materials and $x$. We interpret time-delay changes of the second acoustic peaks $\Delta t$(ii) as the longitudinal speed of sound changes of electrode materials.

The longitudinal speed of sound $v$ was calculated using $v = 2d/(\Delta t_{ii}(x) - \Delta t_i)$ where $d$ is the thickness of the electrode material measured by cross-sectional SEM. We subtract the time for the acoustic wave to travel through the Al film $\Delta t_i = 2d_{Al}/v_{Al}$ where $d_{Al}$ and $v_{Al}$ are the thickness and the longitudinal speed of sound of Al (6.42 nm ps$^{-1}$). For Sb and TiO$_2$, we used the second echoes to calculate $v(x)$, the longitudinal speed of sound of the electrode material at $x$ where the first echoes are week and overlap with echoes from Al/SiO$_2$ interface (dashed lines). Note that we use the theoretical volume expansion ratio of Sb as it is uncertain to define the accurate thickness after lithiation. The measured longitudinal speed of sound $v$ was converted to elastic modulus through the thickness direction $M(x) = \rho(x)v(x)^2$ where $\rho(x)$ is the density of electrode materials at $x$. We measured the mass of $TiO_2$, $V_2O_5$ and $NiO$ at the lithiated and delithiated states using microbalance with an accuracy of 1 µg (XP26, Mettler Toledo) and linearly interpolated them with respect to $x$.

For intercalation electrode materials, we observe lattice stiffening where the echo shifts to shorter delay time with increasing $x$. For conversion and alloying electrode materials, we observe an irreversible lattice softening where the echo shifts to longer delay time with increasing $x$. For alloying material, Sb shows the most dramatic shift in echoes after the lithiation. During delithation, we observe that the echo returned to the initial position due to the reversible phase transition between Sb and $Li_3Sb$ after delithiation (Figure S9d). However, we do not calculate $M$ at the delithiated state, due to the large thickness uncertainty.

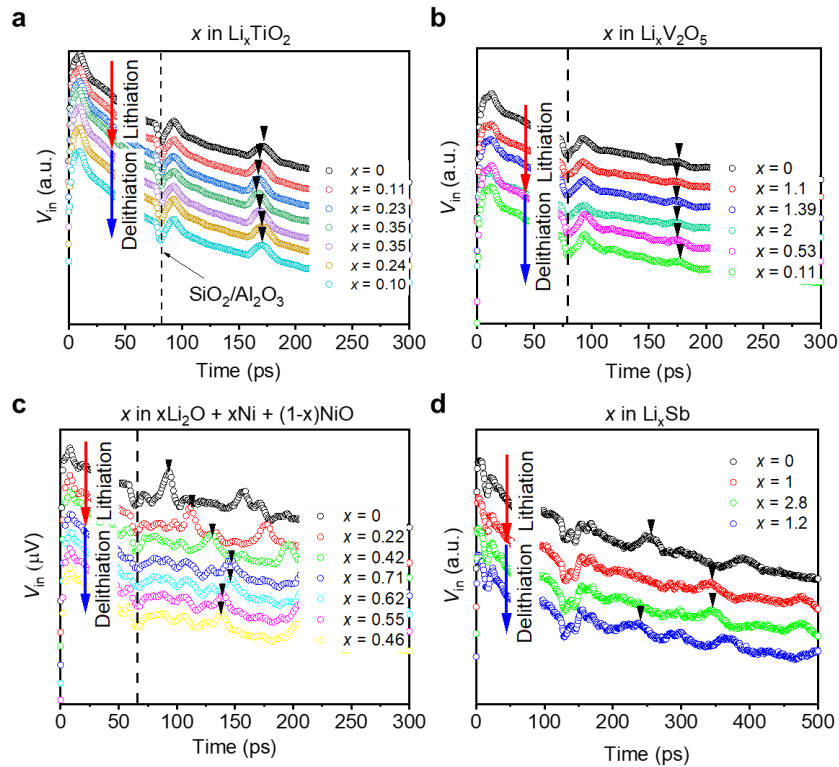

**Figure S9.** Picosecond acoustic signals of electrode materials with varying lithium contents $x$. Measured in-phase TDTR signal of (a) $TiO_2$/Al/$SiO_2$/sapphire, (b) $V_2O_5$/Al/$SiO_2$/sapphire and (c) NiO/Al/$SiO_2$/sapphire in the liquid electrolyte cell. (d) Measured in-phase TDTR signal of Sb/Au/Al/$SiO_2$/sapphire substrate in the solid electrolyte cell.

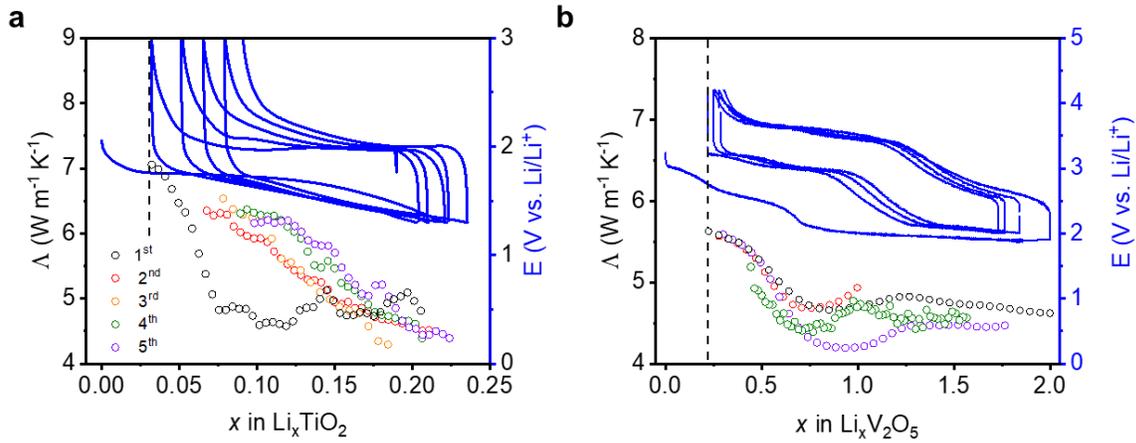

**Figure S10.** $\Lambda$ of intercalation $Li_xTiO_2$ and $Li_xV_2O_5$. (a) $\Lambda$ (delithiation) and potential curves of $Li_xTiO_2$ over five galvanostatic cycles. (b) $\Lambda$ (delithiation) and potential curve of $Li_xV_2O_5$ over five galvanostatic cycles.

**Electrochemical capacity of electrodes**

*Intercalation electrodes*: Figure S11 shows the measured specific charge and discharge capacities ($Q$) of TiO$_2$ and V$_2$O$_5$ during five galvanostatic cyclings presented in the main text. Note that we pre-cycled TiO$_2$ with the initial discharge $Q$ = 110 mA h g$^{-1}$. After pre-cycling, we cycled TiO$_2$ at the current density of 50 mA g$^{-1}$ (3 – 1.3 V vs. Li/Li$^+$) and V$_2$O$_5$ was cycled at 100 mA g$^{-1}$. For 1$^{st}$ cycle, V$_2$O$_5$ was discharged to 1.9 V vs. Li/Li$^+$ (up to $x$ = 2). For the following cycles, V$_2$O$_5$ was discharged to 2.6 V (2$^{nd}$ cycle, $x$ = 1) and 2.0 V vs. Li/Li$^+$ (3$^{rd}$ to 5$^{th}$ cycle, $x \leq$ 2). For all cycles, V$_2$O$_5$ was charged up to 4.2 V vs Li/Li$^+$. The initial discharge $Q$ of both TiO$_2$ and V$_2$O$_5$ were lower than the theoretical capacities of Li$_x$TiO$_2$ (330 mA h g$^{-1}$, $x$ = 1) and Li$_x$V$_2$O$_5$ (260 mA h g$^{-1}$, $x$ = 2) which can be attributed to the poor kinetics of lithium ions diffusion in the thick film (> 200 nm) at high charge/discharge C-rates (1 – 2 C). Here, TiO$_2$ (255 nm) showed discharge capacity fading from 110 mA h g$^{-1}$ (pre-cycling) to 52 mA h g$^{-1}$ (5$^{th}$ cycle) at the current density of 50 mA h g$^{-1}$, close to the reported capacity fading of 300-nm TiO$_2$ microspheres (110 mA h g$^{-1}$ to 50 mA h g$^{-1}$ during 20 cycles).[13] Similarly, we observe a discharge capacity fading V$_2$O$_5$ from 55 mA h g$^{-1}$ (1$^{st}$) to 45 mA h g$^{-1}$ (5$^{th}$).

*Conversion electrodes*: Figure S12 shows the measured specific charge and discharge $Q$ of Fe$_2$O$_3$ and NiO over five galvanostatic cycles described in the main text. Fe$_2$O$_3$ was cycled at the current density of 300 mA g$^{-1}$ (2.5 – 0.5 V vs. Li/Li$^+$) and NiO was cycled at the current density of 182 mA g$^{-1}$ (3.0 – 0.5 V vs. Li/Li$^+$). During the first cycle, Fe$_2$O$_3$ exhibits a discharge capacity $Q$ = 1078 mA h g$^{-1}$. This value is close to the theoretical capacity of Fe$_2$O$_3$ $Q_{theory}$ = 1007 mA h g$^{-1}$ ($x$ = 1). However, this value gradually decreased to 390 mA h g$^{-1}$ during the following cycles. Fe$_2$O$_3$ exhibits poor coulombic efficiency (57%, 69%, 84%, etc.) with continuous capacity fading over five cycles (Figure 3b). On the contrary, NiO shows the first discharge capacity of 508 mA h g$^{-1}$, accounting for 71% of the theoretical capacity (718 mA h g$^{-1}$). After the first discharge, two-thirds of the initial capacity of NiO was not recovered. However, the remaining capacity (~180 mA h g$^{-1}$) remained nearly unchanged for the following cycles with coulombic efficiency higher than Fe$_2$O$_3$ (> 95% from 2$^{nd}$ to 5$^{th}$ cycle).

*Alloying electrode*: During the first cycle, Sb exhibits a discharge capacity $Q$ = 636 mA h g$^{-1}$, close to the theoretical capacity of $Q_{theory}$ = 660 mA h g$^{-1}$ ($x$ = 3). However, this value gradually decreased to 450 mA h g$^{-1}$ during the second cycles. Sb exhibits a large irreversible capacity and poor coulombic efficiency. By limiting discharge, electrochemical capacity retention and coulombic efficiency are improved (from 3$^{rd}$ cycle).

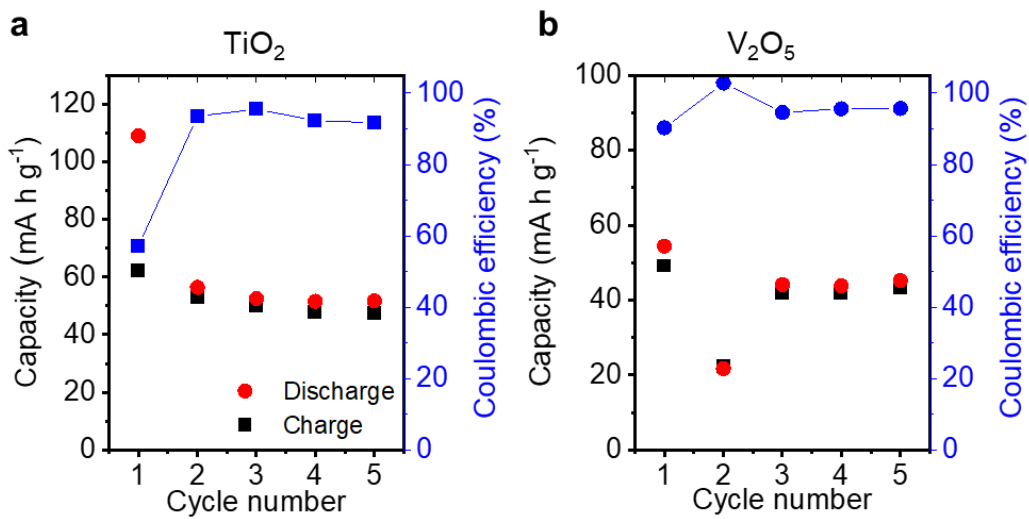

**Figure S11.** Cycling performance of TiO$_2$ and V$_2$O$_5$. Electrochemical charge/discharge capacity of **a** TiO$_2$ and **b** V$_2$O$_5$ during *in situ* TDTR measurements described in the main text. Note that the second cycle of Li$_x$V$_2$O$_5$ was performed with a limiting potential window for the study of thermal conductivity switching over a narrow voltage range. Specific capacities were calculated based on the mass and volume of each electrode material.

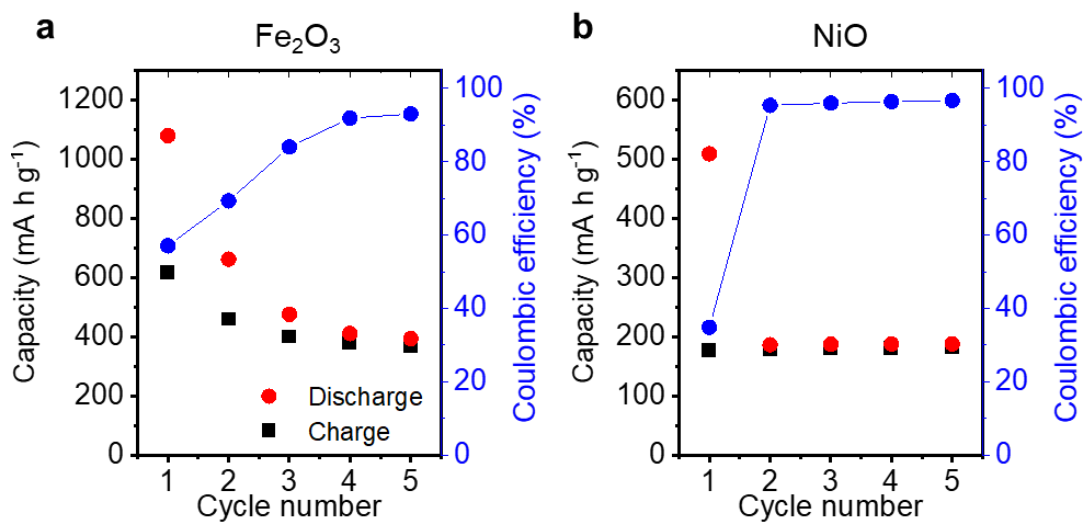

**Figure S12.** Cycling performance of Fe$_2$O$_3$ and NiO. Electrochemical charge/discharge capacity of **a** Fe$_2$O$_3$ and **b** NiO during *in situ* TDTR measurements with electrochemical cycling described in the main text. Specific capacity was calculated based on the mass and volume of each electrode material.